\documentclass[a4paper]{article}

\usepackage[margin=2cm]{geometry}

\usepackage{epstopdf}
\usepackage{subfigure}
\usepackage{tabularx}
\usepackage{booktabs} 

\usepackage{hyperref}
\usepackage{ulem}
\usepackage{amsmath,amsfonts,amssymb}
\usepackage{xcolor}

\usepackage{authblk}
\usepackage{natbib}
\usepackage{tikz}
\usepackage{appendix}

\newtheorem{theorem}{Theorem}[section]

\newtheorem{definition}{Definition}

\newtheorem{remark}{Remark}

\def\E{\mathbb{E}}

\def\Rn{\mathbb{R}^{n}}
\def\Rnp{\mathbb{R}^{n+1}}
\def\Rnn{\mathbb{R}^{n\times n}}
\def\R{\mathbb{R}}
\def\1{\mathds{1}}

\def\S0{S_{\text{init}}}

\def\n_s{n_{\text{s}}}

\def\Xv{\boldsymbol{X}}
\def\xv{\boldsymbol{x}}
\def\wv{\boldsymbol{w}}
\def\vv{\boldsymbol{v}}
\def\rv{\boldsymbol{r}}
\def\uv{\boldsymbol{u}}

\def\bv{\boldsymbol{b}}
\def\yv{\boldsymbol{y}}
\def\zv{\boldsymbol{z}}
\def\qv{\boldsymbol{q}}

\def\lv{\boldsymbol{\lambda}}

\def\mvp{\boldsymbol{m}}
\def\mv{{\boldsymbol{\mu}}}
\def\ev{{\boldsymbol{\epsilon}}}
\def\tv{{\boldsymbol{\theta}}}
\def\Tv{{\boldsymbol{\Theta}}}
\def\Mtwo{\mathbf{M}_2}
\def\Mt{\mathbf{M}_3}
\def\Mf{\mathbf{M}_4}
\def\Sm{\mathbf{S}}
\def\Km{\mathbf{K}}
\def\In{\mathbf{I}_n}
\def\nw{\nabla_{\wv}}
\def\sp{\text{ }}
\def\1v{\boldsymbol{1}}
\def\0v{\boldsymbol{0}}

\def\T{\top}
\def\W{\mathcal{W}}
\def\I{\mathcal{I}}
\def\J{\mathcal{J}}
\def\Q{\mathcal{Q}}

\def\V{\mathcal{V}}
\def\Tp{\mathcal{T}}
\def\X{\mathcal{X}}
\DeclareMathOperator*{\argmax}{argmax}
\DeclareMathOperator*{\argmin}{argmin}

\def\nl{\nonumber \displaybreak[0] \\}
\def\nlt{ \displaybreak[0] \\}
\usepackage{caption}

\begin{document}


\title{Optimising portfolio diversification and dimensionality}

\author[1]{M. Barkhagen}
\author[2]{B. Fleming}
\author[1]{S. Garc\'{i}a}
\author[1]{J. Gondzio}
\author[1]{J. Kalcsics}
\author[2]{J. Kroeske}
\author[1, 3]{S. Sabanis}
\author[2]{A. Staal}

\affil[1]{\footnotesize School of Mathematics, The University of Edinburgh, UK.}
\affil[2]{\footnotesize Aberdeen Standard Investments, Edinburgh, UK.}
\affil[3]{\footnotesize The Alan Turing Institute, UK.}

\renewcommand\Authands{ and }

\date{\today}

\maketitle

\begin{abstract}
A new framework for portfolio diversification is introduced which goes beyond the classical mean-variance approach and portfolio allocation strategies such as risk parity. It is based on a novel concept called portfolio dimensionality that connects diversification to the non-Gaussianity of portfolio returns and can typically be defined in terms of the ratio of risk measures which are homogenous functions of equal degree. The latter arises naturally due to our requirement that diversification measures should be leverage invariant. We introduce this new framework and argue the benefits relative to existing measures of diversification in the literature, before addressing the question of optimizing diversification or, equivalently, dimensionality.  Maximising portfolio dimensionality leads to highly non-trivial optimization problems with objective functions which are typically non-convex and potentially have multiple local optima. Two complementary global optimization algorithms are thus presented. For problems of moderate size and more akin to asset allocation problems, a deterministic Branch and Bound algorithm is developed, whereas for problems of larger size a stochastic global optimization algorithm based on Gradient Langevin Dynamics is given. We demonstrate analytically and through numerical experiments that the framework reflects the desired properties often discussed in the literature.
\end{abstract}

\section{Introduction}
\label{intro}
Diversification as a concept is as old as investing itself, coming into focus in particular during crisis periods. The global financial crisis of 2008, for example,  induced heavy losses for most asset portfolios held by institutional investors, prompting practitioners to question their portfolio construction methodologies and understanding of the level of diversification that can thus be achieved. This led to increased activity in both academia and the financial industry seeking to develop new portfolio construction techniques with the goal of obtaining a well diversified portfolio. Despite the fact that an overwhelming majority of investors seeks to hold a well diversified portfolio, there is still no agreed upon definition or measure of diversification in the literature or among practitioners. A common understanding is that a diversified portfolio should provide risk dissemination and be protected against large drawdowns. Expressed differently, the risk of the portfolio should not be concentrated to only a few risk factors and the tail risk of the portfolio should be controlled. In the diversification literature, the focus was initially purely on volatility reduction, but this definition cannot lead to a useful measure as hedging reduces volatility but does not lead to a better risk dissemination. Most diversification measures and construction methods in the literature are based on the covariance matrix which can be traced back to the seminal paper on mean-variance optimization by \cite{Mar1952}. Other approaches, with a growing literature used by many asset managers and institutional investors, include the  risk parity approach coined by \cite{Qian2005} (see also e.g. \cite{Qian2011}, \cite{Ron2014} and \cite{Ron2016}) and the most diversified portfolio of \cite{Cho2008} (see also \cite{Cho2013}). Another portfolio diversification measure based on the covariance matrix is introduced in \cite{Meu2009}. There, the asset universe is orthogonalized via Principal Component Analysis of the covariance matrix leading to a new universe consisting of uncorrelated so called principal portfolios. Based on the squared weighted volatilities of the portfolio in the new universe, a portfolio diversification measure based on the dispersion of the squared weighted volatilities is defined. The interpretation of this measure put forth by Meucci is that it represents the effective number of uncorrelated bets that the portfolio is exposed to.

The primary drawback with existing portfolio diversification approaches is that they are based on the distribution of portfolio volatilities or marginal contributions to volatility; however, as far as we are aware, there is not a direct connection from these measures to the distributional or sampling properties of portfolio returns. This in turn leads to a situation where different metrics are shown to behave intuitively in particular cases, while singular counterexamples raise questions about how broadly a technique can be applied. For example, unintuitive behaviour is observed in risk parity portfolios when highly correlated assets are added to the portfolio, leading to over-allocation to them. Unintuitive behaviour of the measure introduced in Meucci (2009) led to the introduction of a new technique in Meucci (2013). For this reason we argue, following \cite{Fle2017}, that it is helpful to augment the covariance based frameworks and connect diversification to additional properties of the distribution of portfolio returns such as higher order moments.

Bringing in higher order moments allows us to move beyond limiting Gaussian assumptions. As a relatively extreme example, consider a two strategy portfolio combining an equity index exposure and a volatility selling strategy on the same index. The volatility of most volatility selling strategies is lower than that of the equity index, while the negative skewness and the kurtosis are more pronounced. A portfolio construction strategy based on volatility would thus put larger weight to the volatility selling strategy in order to decrease volatility risk at the expense of being more exposed to tail risk. Such a portfolio would have suffered heavy losses during the VIX spike on 5 February 2018. On that day the VIX index experienced its largest one-day jump in its 25 year history, rising 20 points from 17.31 from the previous day's close to 37.32 at the end of the trading day. This example highlights why we require our diversification measure to have a direct link to the tail properties of the distribution of portfolio returns. Secondly, we want the measure to be leverage invariant as being 100\% exposed to S\&P~500 is as diversified as being 50\% exposed to S\&P~500 and leaving the rest in cash. Thus, the diversification measure should not be based on portfolio volatility or Expected Shortfall alone. After all, the best strategy to reduce volatility or Expected Shortfall is to have more capital allocated in cash but that does not increase the diversification, it simply represents a reduced exposure to risky assets.

In this paper we introduce a framework that offers a coherent foundation for understanding portfolio diversification by connecting it to the non-Gaussian properties of portfolio returns. The requirement that the diversification measure is leverage invariant naturally leads to measures based on ratios of homogeneous functions of equal degree, such as kurtosis (degree 4) or the square of skewness (degree 6). Even though, for example, the fourth moment and the square of variance are both convex functions, the ratio that yields kurtosis is not necessarily convex. Optimizing ratios of convex functions is a global optimization problem with potentially several local optima which are not equal to the global optimum or optima. We therefore develop two methods, one deterministic and one stochastic, for the global optimization of ratios of convex functions and use them to conduct initial numerical experiments.

\section{Non-Gaussianity as a measure of diversification}
In this section we present a framework which first connects non-Gaussianity and diversification before introducing the notion of portfolio dimensionality. The main goal of the portfolio diversification framework is to manage the distribution of the portfolio returns. One observes in fact that, the distribution of portfolio volatilities or marginal contributions to volatility across assets which make up the portfolio is irrelevant within a mean-variance framework. This is due to the underlying assumption of normally distributed returns. All that matters is portfolio variance.  From this simple observation, we argue that meaningful measures of diversification must be related to additional properties of the distribution of portfolio returns. A natural extension of existing frameworks is to connect the concept of diversification to the non-Gaussian properties of the distribution of portfolio returns. Given that the mean-variance framework assumes Gaussianity, diversification can then be seen as an augmentation which relates to model limitations.

\subsection{A novel approach to portfolio diversification: Dimensionality}

In an ideal world, we propose that one could define \textit{portfolio dimensionality} as the number of equally sized independent return streams in the portfolio. This definition is intuitively related to risk dissemination and, arguments based on the Central Limit Theorem (CLT) imply that adding independent exposures to the portfolio leads to a portfolio whose distribution is closer to the Gaussian distribution and thereby the tail risk is reduced. Obviously, financial markets do not obey the idealized assumptions of independent and identically distributed (i.i.d.) returns of the standard CLT \citep[see][for some examples of the CLT with relaxed assumptions]{Bar2010}. The idea behind our diversification measures is to base them on the degree of non-Gaussianity of the portfolio return distribution. A portfolio with a low degree of non-Gaussianity is a well diversified portfolio, and vice versa. Measuring the degree of non-Gaussianity is directly related to the tail properties of the portfolio and naturally leads to measures which are leverage invariant.  Measuring and optimizing non-Gaussianity have been thoroughly studied in the Independent Component Analysis (ICA) literature, see e.g. \cite{Hyv2000}. A common measure of non-Gaussianity in the ICA literature is kurtosis, and other frequently used measures are based on neg-entropy or Kullback-Leibler divergence. Inspired by the ICA literature, we initially link the notion of a well diversified portfolio to a portfolio with a low kurtosis which implies a low (symmetric) tail risk. Other attractive aspects of using kurtosis are that we see it as a natural extension of a symmetric risk framework and it is also related to the distribution of sample variance. In particular, it is known that the variance of the distribution of sample variance is positively related to kurtosis \citep[see e.g.][]{Vaa1998}. Reducing kurtosis therefore increases confidence in estimates of portfolio variance.

That said, asymmetry in the form of skewness is also of interest to investors where empirical results from the risk premia literature (\cite{Lem2017}) show that maximising the Sharpe ratio of a portfolio is strongly linked to maximizing the negative skewness of portfolio returns. There are several ways to incorporate skewness into a portfolio diversification framework. In \cite{Las2018}, a portfolio risk measure based on exponential R\'{e}nyi entropy is used in order to incorporate higher order moments into the portfolio decision framework. Through a truncated Gram-Charlier expansion of R\'enyi entropy they demonstrate that their portfolio risk measure can be directly expressed as a function of portfolio skewness and kurtosis. Another approach, see e.g. \cite{Jon2006}, relies on a higher order Taylor expansion of the investors utility function, which leads to an expression in terms of the non-standardized portfolio moments. This latter approach suffers from the drawback of optimizing an objective function which is not invariant to leverage. In the following, we offer a general framework which allows us to look at various measures including skewness and kurtosis but for the purposes of these initial numerical investigations we focus on kurtosis.

\subsection{From non-Gaussianity to dimensionality: definition and examples}

With all the above in mind, we proceed with defining a diversification framework which is invariant to leverage and directly linked to the tail properties of the distribution of portfolio returns. It is also flexible enough to allow different objective functions, such as excess kurtosis or the square of skewness, but within a robust setting for measuring, in an appropriate sense, the level of non-Gaussianity of resulting portfolios. Furthermore, in order to have an intuitive interpretation of diversification we link it with the tail risk of an equally weighted reference portfolio of i.i.d. reference assets. This reference portfolio is representative of the given asset universe and we proceed to define the notion of portfolio dimensionality relative to the tail risk of the reference asset.

Let $\mathcal{X}$ denote the set of all random variables with mean zero and appropriately finite $p$-th moments, for suitable $p\ge 2$, and with either Gaussian or skewed and leptokurtic distribution. Such random variables represent the asset returns in a given asset universe under consideration. We define a function, which measures the level of non-Gaussianity and is denoted by $\nu: \mathcal{X} \mapsto \mathbb{R}_+$, such that: (i) $\nu(tX) = \nu(X)$, for any $t>0$ and $X \in \mathcal{X}$ (leverage invariant), and (ii) the function
\begin{equation} \label{f}
f_{Y,\nu}(n)= \nu(\sum_{i=1}^{n}Y_i), \qquad \mbox{where } Y_i \in \mathcal{X} \mbox{ are i.i.d. and } n\in\mathbb{N},
\end{equation}
is strictly decreasing in $n$ and $\rm{Law}(Y_1)= \rm{Law}(Y)$. Then, one proceeds with defining a measure of diversification relative to a reference random variable $Z$, as a continuous function

\begin{flalign} \label{portfolio-div}
\mbox{(\textbf{Diversification measure}):} \qquad  D_{Z,\,\nu}(w) = \frac{\nu(Z)}{\nu\left(\sum_{i=1}^{n}w_iX_i\right)}, &&
\end{flalign}
where $X_i$ is the return of the $i$-th asset of the portfolio and $w_i$ is the corresponding weight. Moreover, due to the leverage invariance of $\nu$ and the strict monotonicity of the function $f$, see \eqref{f}, one defines a function $h:\mathbb{R}_+ \mapsto \mathbb{R}_+ $, which is the continuous, monotonic interpolation of
\[
\hat{h}(k): = \frac{\nu(Z)}{\nu\left(\frac{1}{k}\sum_{i=1}^{k}Z_i\right)}, \mbox{ for any }  k\in\mathbb{N},
\]
and which is a strictly increasing function. Hence, the definition of portfolio dimensionality follows naturally by considering the following transformation
\begin{flalign} \label{portfolio-dim}
\mbox{(\textbf{Portfolio dimensionality}):} \qquad  d_{Z,\,\nu}(w) = h^{-1}\left(D_{Z,\,\nu}(w)\right). &&
\end{flalign}

\begin{remark}
One observes that, for any $k\le n$,
\begin{equation} \label{portfolio-dim-calc}
D_{Z,\,\nu}(w) = \frac{\nu(Z)}{\nu\left(\frac{1}{k}\sum_{i=1}^{k}Z_i\right)}\frac{\nu\left(\sum_{i=1}^{k}Z_i\right)}{\nu\left(\sum_{i=1}^{n}w_iX_i\right)} = h(k) \frac{\nu\left(\sum_{i=1}^{k}Z_i\right)}{\nu\left(\sum_{i=1}^{n}w_iX_i\right)},
\end{equation}
where $\{Z_i\}_{1\le i\le n}$ is a sequence of i.i.d. random variables such that $\rm{Law}(Z_1)= \rm{Law}(Z)$. Thus,
\[
\nu\left(\sum_{i=1}^{n}w_iX_i\right) = \nu\left(\frac{1}{k}\sum_{i=1}^{k}Z_i\right) \qquad \Rightarrow \qquad  d_{Z,\,\nu}(w)=k,
\]
since the denominator in \eqref{portfolio-dim-calc} is equal to $f_{Z,\nu}(k)$, which implies that $D_{Z,\,\nu}(w)= h(k)$. Thus, we see that the portfolio dimensionality is exactly the number of independent return streams.

\end{remark}

\begin{remark}
In general, a judicious selection of the reference asset as representative of the investment universe will produce values of $D_{Z,\,\nu} > 1$ as we achieve some relative diversification benefit; however, we note that $D_{Z,\,\nu}<1$ is also possible if we worsen the relative tail properties.
\end{remark}

Let us concentrate now on the case where $\nu$ is either excess kurtosis or the square of skewness. Using the leverage invariance property of $\nu$ and taking into account the findings of \cite{Fle2017}, where the notion of the distribution of portfolio variance is used and the effective size of its support is related to the spectrum of Rényi entropies, one identifies $h$ with the identity function and $d_{Z,\,\nu}(w) = D_{Z,\,\nu}(w)$. Thus, we see that the diversification measure under either excess kurtosis or the square of skewness satisfies our definition of portfolio dimensionality. Moreover, since $f_{Z,\nu}$ is a monotonically decreasing function, if
\[
f_{Z,\nu}(k+1) <  \nu\left(\sum_{i=1}^{n}w_iX_i\right) < f_{Z,\nu}(k) \mbox{, then } k < D_{Z,\,\nu}(w) < k +1,
\]
with $D_{Z,\,\nu}(w)$ taking a non-integer value according to a monotonic interpolation of $f_{Z,\nu}$. As a result, one observes that the higher the values for  $D_{Z,\,\nu}(w)$, the closer we are to a tail risk similar to the one given by a standard Gaussian. To see this, consider a large enough $k$ and $ D_{Z,\,\nu}(w) \ge k $. Consequently, one obtains due to \eqref{portfolio-dim-calc}, that the number of independent assets is increased accordingly,
\[
\nu\left(\sum_{i=1}^{n}w_iX_i\right) \le f_{Z,\nu}(k) = \nu\left(\frac{\sum_{i=1}^{k}Z_i}{\sqrt{k\mathbb{E}[Z^2]}}\right), \qquad \mbox{(due to leverage invariance)}
\]
and thus due to the CLT and property (ii), one observes the desired result.

\subsection{Desirable properties of the diversification measure: Toy example}
Although there is no agreed upon definition of diversification in the literature, a number of desirable properties of a diversification methodology have been proposed. In \cite{Cho2013}, the notion of \textit{polico invariance} is introduced. Extending an asset universe by adding a positive linear combination of assets already belonging to the universe should not affect the weights to the original assets when applying the diversification methodology. A special case of polico invariance, denoted duplication invariance, considers the duplication of one of the assets in the universe. This case naturally arises in applications when one of the assets is listed on multiple exchanges. Applying the diversification methodology should produce the same portfolio irrespective of any asset in the universe being duplicated or not. In \cite{Kou2017} further desirable properties of diversification measures are introduced. However, some of the properties presented in \cite{Kou2017} are not consistent with the requirements that we have on a diversification measure. In Section \ref{intro}, we introduced the requirement that the portfolio diversification measure should be leverage invariant. This contrasts one of the desired properties presented in \cite{Kou2017}. Furthermore, in \cite{Kou2017}, the portfolio diversification measure is required to be concave or quasi-concave. As we have argued, a leverage invariant diversification measure typically leads to a ratio of two convex functions which in general is neither concave nor quasi-concave.

In the following, a numerical example is used to demonstrate that important desirable properties are satisfied by the newly introduced portfolio diversification measure. The demonstration is based on a toy example with a universe consisting of three assets with the following covariance matrix
\begin{align}
  \mathbf{C}= \left[
\begin{array}{c c c }
1 \sp &  \rho \sp & 0 \\
\rho \sp & 1 \sp & 0 \\
0 \sp & 0 \sp & 1
\end{array}
\right].
\end{align}
As the correlation $\rho$ between asset one and asset two approaches one, these two assets behave as one asset and hence this corresponds to the case when one of the assets in the universe is duplicated. For this case, the weight of asset three should approach $\tfrac{1}{2}$ as $\rho \rightarrow 1$. When $\rho \rightarrow -1$, this corresponds to the case when either asset one or asset two is a perfect hedge of the other. In this case, assuming that $\mathbf{C}$ is positive definite, the volatility of a portfolio given by the weight vector $\wv=[0.5, \sp 0.5, \sp 0]^{\T}$ tends to a small value $c>0$ as $\rho\rightarrow -1$. In \cite{Cho2013}, it is demonstrated that risk parity suffers from duplication invariance. It is well known in the literature that the global minimum variance portfolio tends to be highly concentrated to assets with low volatility, see e.g. \cite{Cla2013}. Thus, for an asset universe where the exposure to some assets to a large extent has been hedged away, the global minimum variance portfolio tends to be highly concentrated to the hedged assets. We denote this undesirable property of the global minimum variance portfolio the hedging invariance problem.

\begin{figure}
\begin{center}
\subfigure[]{
\resizebox*{.48\linewidth}{!}{\includegraphics{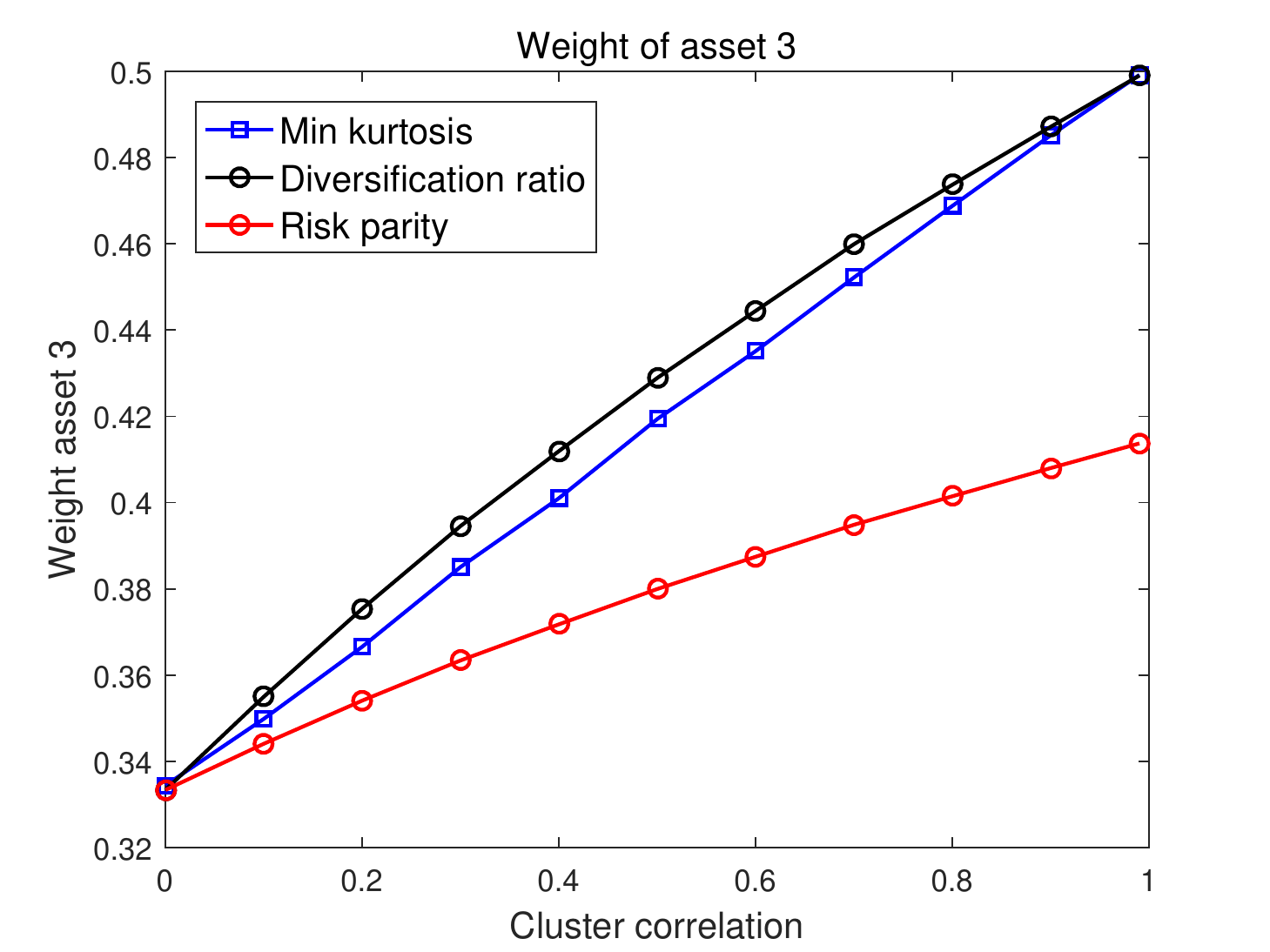}}\label{posCorr}}
\subfigure[]{
\resizebox*{.48\linewidth}{!}{\includegraphics{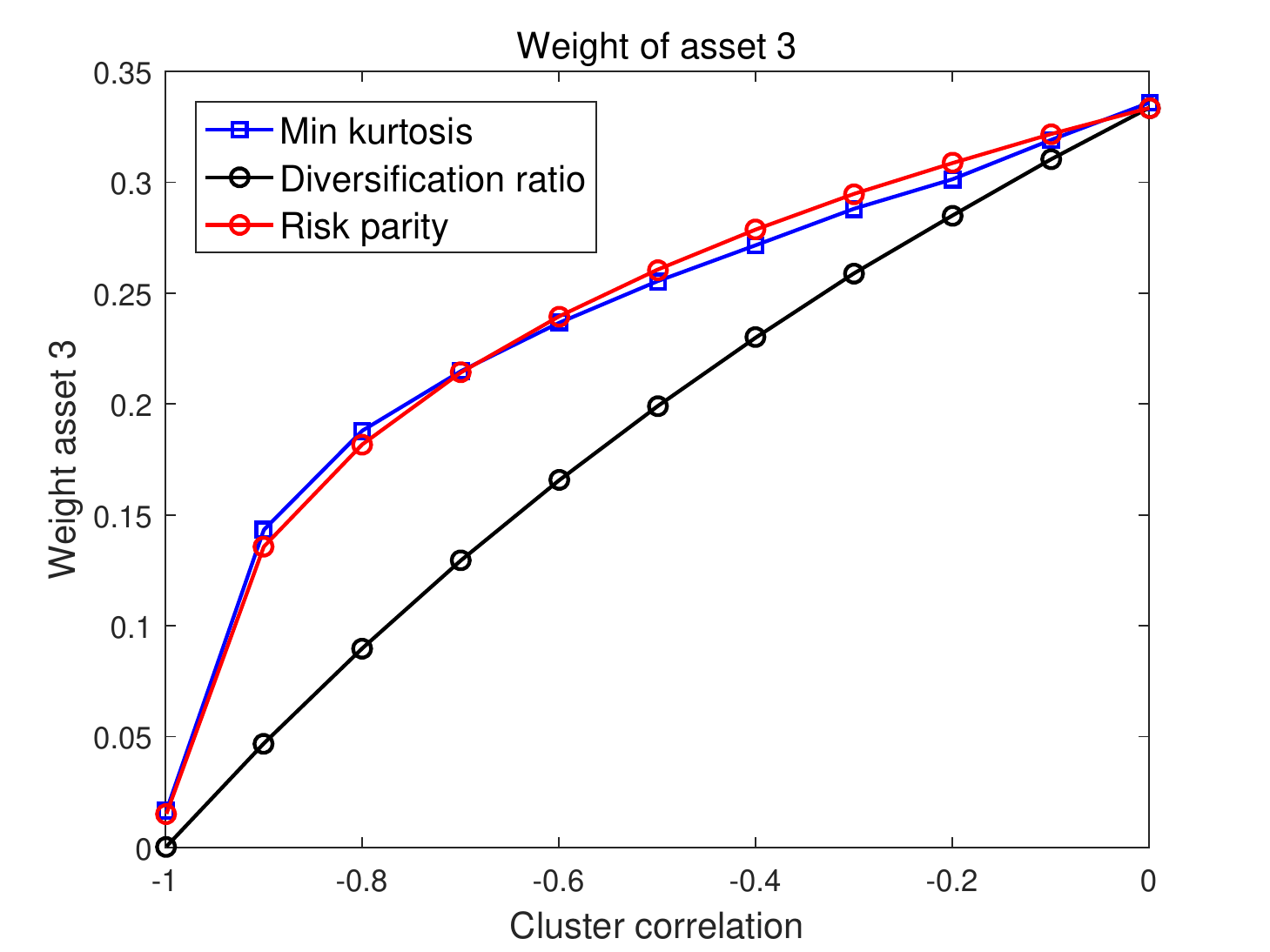}}\label{negCorr}}
\captionsetup{font=small}
\caption{Weight of asset three for the minimum kurtosis, risk parity and maximized diversification ratio portfolios for the cases when: (a) $\rho\in[0,1]$ and (b) $\rho \in (-1,0]$.}
\label{duplHedge}
\end{center}
\end{figure}

Consistency with the duplication invariance and hedging invariance properties for the introduced diversification framework is illustrated in Figure \ref{duplHedge} for the case when the marginal distributions of the assets can be assumed to be approximately symmetric. In this case, we assume that non-Gaussianity is adequately captured by portfolio kurtosis. The consistency with the desired properties is monitored through the weight of asset three for the cases when $\rho \rightarrow 1$ and $\rho \rightarrow -1$. The weight of asset three obtained when minimizing portfolio kurtosis is compared to the corresponding weights obtained with risk parity and from maximizing the diversification ratio introduced in \cite{Cho2008}. Since the volatilities of the three assets are equal, the portfolio obtained from maximizing the diversification ratio coincides with the global minimum variance portfolio, see \cite{Cho2008}. For risk parity and the most diversified portfolio, the weight of asset three can be solved analytically and is given by
\begin{align}
  w_3^{\text{RP}}=\dfrac{2\sqrt{1+\rho}-(1+\rho)}{3-\rho},
\end{align}
for the risk parity portfolio, and
\begin{align}
  w_3^{\text{DR}}=\dfrac{1+\rho}{3+\rho},
\end{align}
for the maximized diversification ratio and the global minimum variance portfolios. Thus, when $\rho \rightarrow 1$, the weight of the third asset approaches $\sqrt{2}-1$ for the risk parity portfolio, whereas $w_3 \rightarrow \tfrac{1}{2}$ for the maximized diversification ratio. From Figure \ref{posCorr}, one observes that the minimum kurtosis portfolio and the maximized diversification ratio satisfy the duplication invariance property, whereas risk parity does not.

When $\rho \rightarrow -1$, the volatility of a portfolio with weight vector $\wv =[0.5, \sp 0.5, \sp 0]^{\T}$ approaches the small value $c>0$. All portfolio construction methodologies that are based on only the covariance matrix will approach this solution when $\rho \rightarrow -1$. The question is at which rate. From Figure \ref{negCorr}, one observes that $w_3^{\text{DR}}$ approaches zero at a rate which is close to linear when $\rho$ varies between 0 and -1. Since this corresponds to the behaviour of the global minimum variance portfolio, which suffers from the hedging invariance problem, this rate is too large when $\rho$ is not close to -1. Figure \ref{negCorr} reveals that the weight of asset three for both the minimum kurtosis and risk parity portfolios approaches zero at a slower rate compared to the diversification ratio portfolio when $\rho$ is not close to -1. These portfolios are thus not too heavily concentrated to the partially hedged exposure represented by asset one and asset two in our example. We conclude that the minimum kurtosis and risk parity portfolios satisfy the hedging invariance property. Hence, only the minimum kurtosis portfolio satisfies the two desired properties when the asset distributions are symmetric.

We finally stress that in this paper we do not attempt to accurately estimate higher order moments or joint distributions of assets returns. The multivariate distribution of the asset returns is modelled with a Gaussian copula and marginal distributions which allow for differing skewness and kurtosis parameters for the individual assets. By modelling the dependence structure with a Gaussian copula, we avoid the notoriously difficult task of estimating a nonlinear dependence structure between the assets. The cost of using a model with less uncertainty in the estimated parameters is that we only take linear dependence between the assets returns into account in this paper. In order to obtain a robust implementation of the framework we take the approach of assigning representative tail risk parameters for different asset groups. Based on the assigned tail risk parameters, the diversification framework then lets us measure and optimize the portfolio dimensionality for a given asset universe.

\section{Deterministic global optimization of ratios of convex functions}

There are numerous applications in finance that involve the optimization of ratios, see, e.g., \cite{Stoy2007}. In the previous two sections we argued that formulating an appropriately defined portfolio diversification measure naturally leads to functions that are ratios of convex functions. In this section we develop a deterministic algorithm for solving such problems to global optimality.

Let $\X \subseteq \R^n$ be a nonempty compact convex set and consider the maximization problem
\begin{align}
 \max\limits_{\xv \in \X}\, h(x), \qquad\text{where}\qquad h(x) \:=\: \dfrac{f(x)}{g(x)}
\end{align}
and $f,g: \X \rightarrow \R$ are positive and continuous functions. In \cite{Avr1988} it is shown that when $f$ is concave and $g$ is convex, then $h(x)$ is a strictly quasi-concave function. Many theoretical results, as well as algorithms of convex programming, apply to the problem of maximizing a strictly quasi-concave function over a convex set \citep[see][]{Dink1967,Schaib1974,Schaib1976}. In particular, each local maximum is again a global maximum.
For the case when $f$ and $g$ are either both convex or both concave, $h(x)$ is in general neither a quasi-concave nor quasi-convex function and the function may have multiple local optima that are different from the global optimum.

\subsection{Formulation of the portfolio kurtosis minimization problem}

Using the notation for higher order portfolio moments introduced in Appendix \ref{app1}, the portfolio kurtosis as a function of the portfolio weights can be expressed as
\begin{align}\label{kurt_p}
  \kappa_p(\wv) \:=\: \dfrac{\E\left((\wv^{\T}(\rv-\mv))^4\right)}{(\E\left(\left(\wv^{\T}(\rv-\mv))^2\right)\right)^2} \:=\: \dfrac{\wv^{\top}\Mf(\wv\otimes\wv\otimes\wv)}{\left(\wv^{\T}\Mtwo \wv \right)^2},
\end{align}
where $\wv \in \Rn$ is the vector of relative portfolio weights, $\rv \in \Rn$ denotes the vector of asset returns, $\mv=\E(\rv)$, and $\Mtwo \in \Rnn$ and $\Mf \in \R^{n\times n^3}$ denote the covariance and fourth co-moment matrices of the asset returns, respectively. We assume that $\Mtwo$ is positive definite and hence that $\wv^{\T}\Mtwo \wv>0$ for all non-zero $\wv$. Therefore, the ratio (\ref{kurt_p}) is well defined. By application of Jensen's inequality we also have that
\begin{align}
  \E\left((\wv^{\T}(\rv-\mv))^4\right)\geq (\E\left(\left(\wv^{\T}(\rv-\mv))^2\right)\right)^2
  \qquad\text{and thus}\qquad
  \wv^{\top}\Mf(\wv\otimes\wv\otimes\wv)>0 \,.
\end{align}
The convention for the majority of papers in the fractional programming literature is to formulate the fractional program as a maximization problem. Since $\argmax_{\wv} f(\wv)/g(\wv) = \argmin_{\wv} g(\wv)/f(\wv)$, for $f(\wv)>0$ and $g(\wv)>0$, we formulate the portfolio kurtosis optimization problem as the following maximization problem, which we denote by (P)
\begin{align}\label{optdef}
  (\text{P})  & \hspace{0.5cm} \max\limits_{\wv \in \mathcal{W}}h(\wv), \qquad\text{where}\qquad
  h(\wv)=\dfrac{f(\wv)}{g(\wv)}=\dfrac{\left(\wv^{\T}\Mtwo \wv \right)^2}{\wv^{\top}\Mf(\wv\otimes\wv\otimes\wv)}
\end{align}
and $\mathcal{W}$ denotes the feasible set for the weights. Since we assume no short selling and a fully invested portfolio, the feasible set is given by $\W=\{\wv \in \Rnp \sp \mid \sum_{i=1}^{n} w_i = 1, w_i \ge 0, i=0,\ldots,n \}$. Letting $\wv^*$ denote the optimal weights, the minimum kurtosis over the feasible set is then given by $\kappa_p(\wv^*)=1\big/ h(\wv^*)$. Since $\Mtwo$ is positive definite, the numerator in (\ref{optdef}) is a convex function. In \cite{Ath2003} it is shown that the fourth moment of the portfolio return is a convex function and hence \eqref{optdef} is a ratio of two convex functions.

\subsection{Branch~and~Bound algorithm for global minimization of portfolio kurtosis}\label{BB}

Global optimization of ratios of convex functions is a very difficult optimization problem and has attracted attention in the optimization research community. In this section we present a Branch~and~Bound (BB) algorithm for global minimization of portfolio kurtosis.
The basic idea of BB is to recursively subdivide the solution space geometrically into smaller and smaller subsets, until we can either compute the optimal solution over a subset or rule out that a subset contains the global optimum. A crucial component of the algorithm and key to its efficiency, is the derivation of tight upper and lower bounds on the objective function value, both globally and locally for each subset.
Examples of papers in the literature which develop BB algorithms for the special case of ratios of convex quadratic functions are \cite{Got2001}, \cite{Ben2006b} and \cite{Yam2007}. The first, and to the best of our knowledge only, paper which develops a BB algorithm for global optimization of a single ratio of general convex functions is \cite{Ben2006}. The generalized problem of optimizing a sum of ratios of convex functions has also attracted considerable attention in the literature. In \cite{Shen2013} a BB algorithm for global optimization for the sum of ratios of convex functions over a convex set is developed, while \cite{Shen2009} develop a BB algorithm for the case of optimizing the sum of ratios of convex functions when the feasible set is non-convex. Comprehensive treatments of BB algorithms for global optimization can be found in \cite{Horst1996} and \cite{Flou2000}.

We apply the BB algorithm developed by \cite{Ben2006} to the problem of portfolio kurtosis minimization and improve the convergence rate by constructing considerably tighter bounds.
In the following we first give an overview of the BB algorithm before we describe the steps of the procedure in more detail.
As input to the algorithm, one chooses an error tolerance $\rho$ which determines the maximum allowed relative distance between the output value of the algorithm and the global optimum. The output of the algorithm is a $\rho$-globally optimal solution:
\begin{definition}[$\rho$-globally optimal solution]
A solution $\wv^k\in \W$ for problem (P) is called $\rho$-globally optimal, if $h(\wv^k) \geq (1-\rho)h(\wv^*)$, where $\rho \in [0,1)$ and $\wv^*$ is an optimal solution for (P).
\end{definition}

\noindent
The basic idea of the BB algorithm is rather simple and consists of the following elements.

\paragraph{Branching process}
Consists of choosing a subset $S \subseteq \W$ that is to be subdivided, and then applying a partitioning method for splitting this subset into two smaller subsets.

\paragraph{Upper bounding process}
Consists of solving a subproblem to obtain an upper bound $UB(S)$ for the maximum of $h(\wv)$ over each subset $S \subseteq \W$ created by the branching process.
Moreover, the upper bound for each subset is used to update a global upper bound $UB$ for the maximum of $h(\wv)$ over $\W$.

\paragraph{Lower bounding process}
Consists of calculating a lower bound $LB(S)$ for the maximum of $h(\wv)$ over each subset $S \subseteq \W$ created by the branching process.
Moreover, the lower bound for each subset is used to update the global lower bound $LB$ for the maximum of $h(\wv)$ over $\W$.

\paragraph{Fathoming process} Deletes each subset $S \subset \W$ in the partition which satisfies $(1-\rho)UB(S) \leq LB$. The algorithm stops when all subsets have been fathomed, i.e., the partition is empty.
\medskip

Unlike heuristic methods, BB algorithms terminate with the guarantee that the value of the best found solution is $\rho$-globally optimal. BB algorithms are however often slow, and in many cases they require computational effort that grows exponentially with the problem size. This is due to the fact that the size of the partition will grow from iteration to iteration, unless we can fathom subsets. Fathoming subsets, however, depends on the quality of the lower and, especially, the upper bound for a subset.
If the upper bound is loose, then a good feasible solution found early in the search may be detected as good only much later in the partitioning process. In other words, the main computational burden of the BB algorithm typically comes from proving global optimality of a feasible point found at an early stage.
Thus, in order for the BB algorithm to be efficient, it is crucial to carefully model the functions used for producing upper bounds for each subset generated by the branching process, to be able to fathom them as quickly as possible. Compared to the BB algorithm in \cite{Ben2006}, we develop two extensions which provide much tighter upper bounds and, thereby, considerably speed up the convergence of the algorithm.
Next, we will give a more detailed description of the BB algorithm applied to the problem of minimizing portfolio kurtosis.

\subsubsection{Branching process}

The branching process splits the feasible set into successively finer partitions.
We denote by $\Q_0=\{\W\}$ the initial partition and by $\Q_k=\{S_i\}_{i \in \I_k}$ the partition in iteration $k$ of the BB algorithm, where $\I_k$ is a finite index set, $\W = \bigcup_{i\in \I_k} S_i$, and $int(S_i) \cap int(S_j) = \emptyset$, for $i \neq j$.
Note that, strictly speaking, once we start fathoming subsets, $\Q_k$ will no longer form a partition of $\W$. However, for the ease of exposition, we will still call $\Q_k$ a partition.
At the beginning of step $k\geq1$, the partition $\Q_{k-1}$ consists of subsets not yet deleted by the algorithm. To determine the subset of $\Q_{k-1}$ to be partitioned, we follow the classical \emph{best-first} rule, which selects the subset $S^k \in \Q_{k-1}$ with the largest upper bound. The rationale for this rule is to pick a subset which is likely to contain a good feasible solution, which will, hopefully, allow for a quick increase in the global lower bound and thereby speed up the fathoming process. See \cite{Loc2013} for other common rules.

First, we observe that our feasible set $\W$ is identical to the standard $n$-simplex.
In order to refine a partition $\Q_{k-1}$, we follow \cite{Ben2006} and split the chosen subset $S^k$ into two halves by simplicial bisection, which is a special case of radial subdivision introduced in \cite{Horst1976}:
\begin{definition}[Radial subdivision]\label{radsub}
  Let $M$ be an $n$-simplex with vertex set $\V(M)=\{\vv^0, \vv^1, \hdots, \vv^n  \}$. Choose a point $\mvp \in M, \mvp \notin \V(M)$ which is uniquely represented by
  \begin{align*}
  \mvp=\sum\limits_{i=0}^n \lambda_i \vv^i, \sp \lambda_i \geq 0 \sp (i=0,\hdots,n), \sp \sum\limits_{i=0}^n \lambda_i =1,
  \end{align*}
  and for each $i$ such that $\lambda_i>0$ form the simplex $M(i,\mvp)$ obtained from $M$ by replacing the vertex $\vv^i$ by~$\mvp$, i.e., $M(i,\mvp)=\{\vv^0, \hdots, \vv^{i-1}, \mvp, \vv^{i+1}, \hdots, \vv^n\}$.
\end{definition}

\begin{figure}
\begin{center}
\subfigure[]{
\includegraphics[width=.31\linewidth]{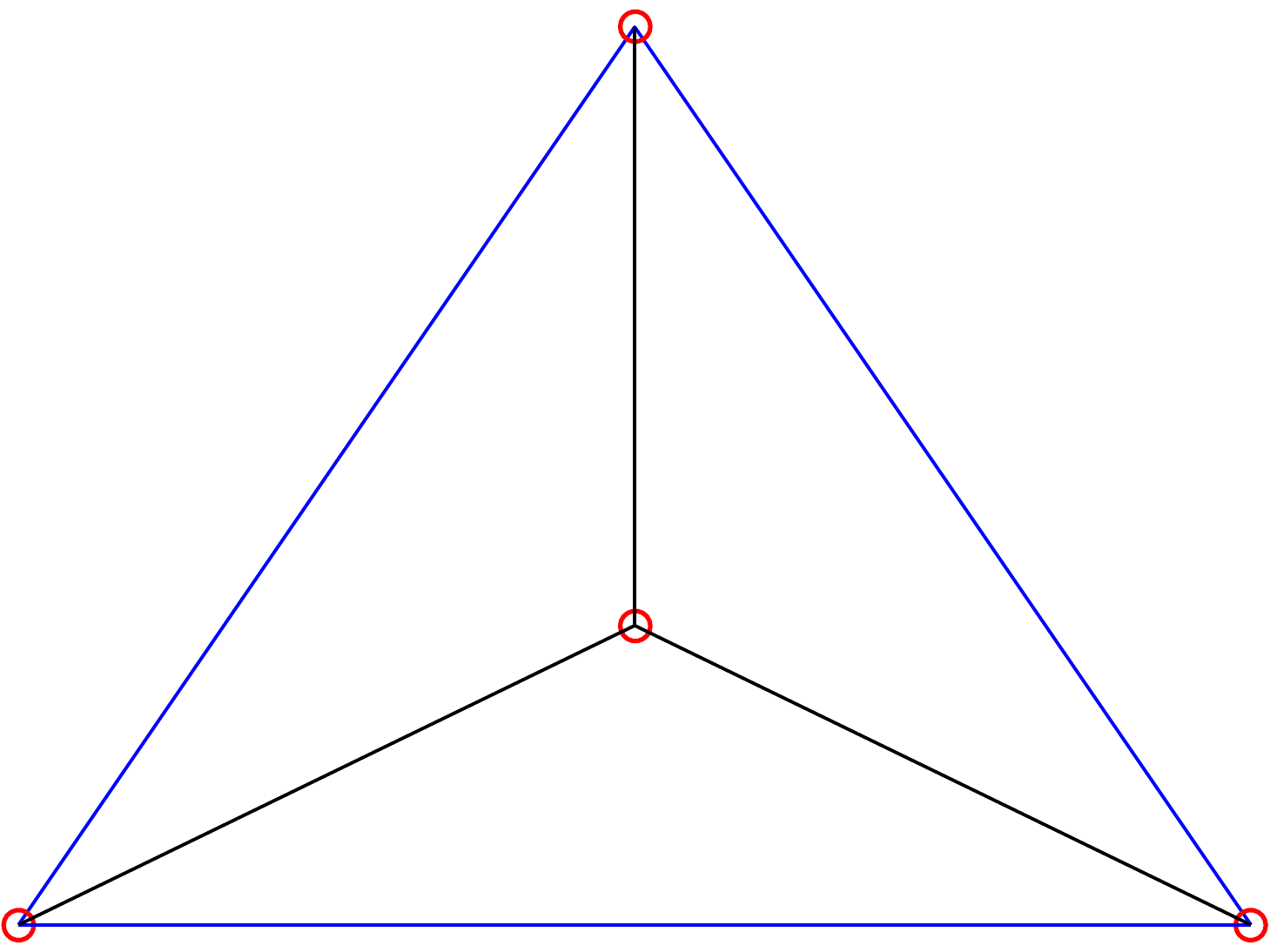}}
\hspace*{2cm}
\subfigure[]{
\includegraphics[width=.31\linewidth]{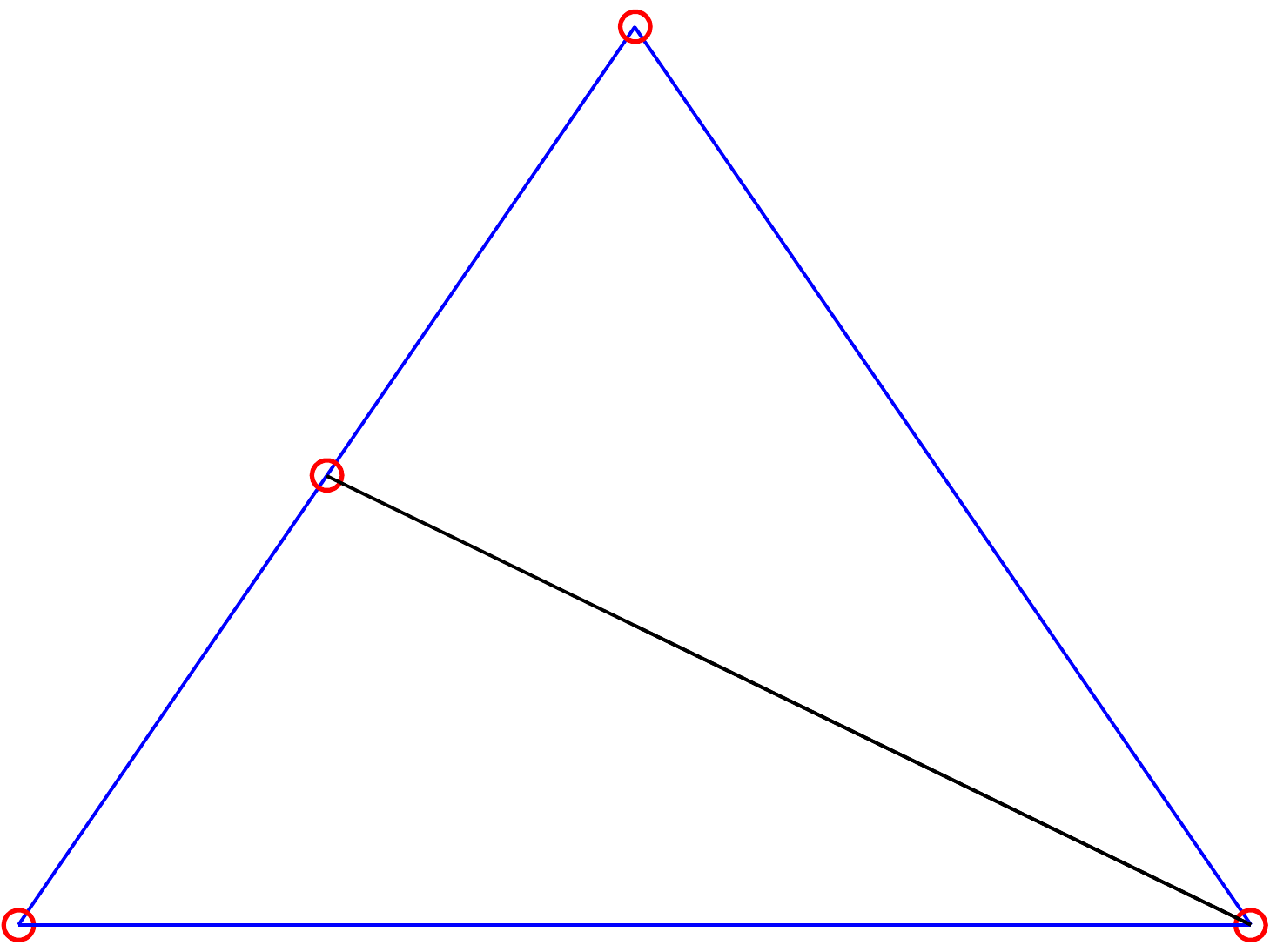}}
\captionsetup{font=small}
\caption{Examples of subdivision of a 2-simplex: radial subdivison (a) and simplicial bisection (b).}
\label{subdivfig}
\end{center}
\end{figure}

A simplicial bisection is obtained by choosing $\mvp$ as the midpoint of a longest edge of the simplex $M$, see Figure~\ref{subdivfig} for an example. \cite{Horst1996} prove that the set of subsets $M(i,\mvp)$ that can be constructed from an $n$-simplex $M$ by an arbitrary radial subdivision forms a partition of $M$ into $n$-simplices. Hence, our subsets $S_i$ are again $n$-simplices. Let $\hat{\vv}$ denote the midpoint of one of the longest edges of $S^k$ and $\vv^d$, $\vv^e$ the corresponding endpoints of this edge. In the branching process, we replace $S^k$ by the two $n$-simplices with vertex sets $S_1^k= M(d,\hat{\vv})$ and $S_1^k= M(e, \hat{\vv})$ using simplicial bisection to obtain a refined partition $\Q_k = (\Q_{k-1}\setminus\{S^k\}) \cup \{S_1^k,S_2^k \}$.

\subsubsection{Upper bounding process}

Let $S \in \Q_k$ be an $n$-simplex of the partition with vertices $\{\vv^0, \vv^1, \hdots, \vv^n \}$. 
Initially, we follow \cite{Ben2006} and overestimate the objective function $h(\wv)=f(\wv)/g(\wv)$ by the ratio of two affine functions: one that overestimates $f$ and one that underestimates $g$. We will improve these bounding functions in Section~\ref{sec:BB:improved_upper_bound} in order to obtain tighter upper bounds and thereby increase the speed of convergence. The function $g$ in the denominator is underestimated by a first order Taylor expansion around the barycenter $\hat{\vv}=1/(n+1)\sum_{i=0}^n \vv^i$ of the simplex $S$ according to
\begin{align}
  g_S(\wv)=g(\hat{\vv})+\nw g(\hat{\vv})(\wv-\hat{\vv})^{\T}\,.
\end{align}
As $g$ is a convex function, $g_S(\wv) \leq g(\wv)$, $\wv \in \Rnp$, and, hence, $g_S$ is an underestimator of $g$. The gradient of the fourth central moment of the portfolio return is given by (see Appendix \ref{app1})
\begin{align}
  \nw g(\wv)=4\Mf(\wv\otimes\wv\otimes\wv).
\end{align}
In order to ensure that the approximation is positive, let
\begin{align}\label{zeq}
  z(\wv)\:=\:\max \{\alpha, g_S(\wv) \}, \qquad\text{where}\quad \alpha\,=\,\min\limits_{\wv \in \W} g(\wv)\,.
\end{align}
With $g$ being convex, the minimization problem on the right-hand side can be solved efficiently.

In order to construct a linear overestimator of the function $f$ in the numerator we need the following definition given in \cite{Horst1996}:
\begin{definition}[Concave envelope]
  The concave envelope of a function $p$ taken over a nonempty subset~$M$ of its domain is the function $p^M$ that satisfies:
  \begin{align*}
  (i) \sp & p^M \text{ is a concave function defined over the convex hull of } M, \nl
  (ii) \sp & p^M(\xv) \geq p(\xv), \text{ for all } \xv \in M, \text{ and} \nl
  (iii) \sp & \text{if } q \text{ is a concave function defined over the convex hull of } M \text{ that satisfies } q(\xv) \geq p(\xv) \\[-1ex]
            &\text{for all } \xv \in M, \text{ then } q(\xv) \geq p^M(\xv)  \text{ for all } \xv \text{ in the convex hull of } M.
\end{align*}
\end{definition}

\cite{Horst1976} shows that when $M$ is an $n$-simplex and $p$ is a convex function on $M$, then $p^M$ is the unique affine function that coincides with $p$ at the vertices of $M$.
Denoting by $f^S(\wv)$ the concave envelope of $f$ over $S$, we construct the following upper bound for the maximum of $h$ over $S$
\begin{align}
  UB(S)=\max\limits_{\wv \in S} \dfrac{f^S(\wv)}{z(\wv)}.
\end{align}
Since $z(\wv)\geq 0$, $\wv \in \Rnp$, and $f^S(\wv)\geq f(\wv) > 0$, $\wv \in S$, $UB(S)$ is equal to the optimal value of the following problem:
\begin{align}
\text{(P1($S$))} \hspace{0.7cm} \max\limits_{t,\wv \in  S} \quad & \dfrac{f^S(\wv)}{t} \notag\\
\text{s.t.} \quad & t\geq \alpha, \label{P1:t_alpha} \\
& t-g_S(\wv)\geq 0. \label{P1:t_gS}
\end{align}
As $S \subseteq \W$ is compact and the objective function is continuous, (P1($S$)) has an optimal solution. Moreover, as the ratio of two linear functions is quasi-concave, every local optimum over the closed convex set is also a global optimum. Thus, the fractional program can be solved to global optimality with any local solver.
However, as P1($S$) has to be solved many times during the BB algorithm, we follow \cite{Ben2006} and reformulate the problem as follows.
Each $\wv \in S$ can be written as
\begin{align}\label{convcomb1}
  \wv \:=\: \sum_{i=0}^n \lambda_i \vv^i, \quad\text{where}\quad \lambda_i \geq 0, i=0,1,\ldots,n, \text{ and }\sum_{i=0}^n \lambda_i =1,
\end{align}
\citep[see][]{Horst1996}. As $f^S(\wv)$ is an affine function, we then get
$f^S(\wv) \,=\, \sum_{i=0}^n \lambda_i f(\vv^i)$.
Substituting $f^S(\wv)$ and adding the conditions for $\wv$ gives the equivalent fractional program
\begin{align}
\text{(P2($S$))} \hspace{0.7cm} \max\limits_{t,\lambda,\wv} \quad & \dfrac{1}{t}\sum_{i=0}^n \lambda_i f(\vv^i) \notag \\
\text{s.t.} \quad & \eqref{P1:t_alpha}, \eqref{P1:t_gS}, \notag \\
& \wv = \sum_{i=0}^n \lambda_i \vv^i, \label{P2:w} \\
& \sum_{i=0}^n \lambda_i =1, \label{P2:sum_lambda} \\
& \lambda_i \geq 0, \quad i=0,1,\ldots,n. \label{P2:dom_lambda}
\end{align}

To linearize the objective function, we apply the Charnes-Cooper transformation \citep{Cha1962}, performing the following change of variables
\begin{align}\label{varchange}
  u=\dfrac{1}{t}, \quad b_i=\dfrac{\lambda_i}{t}, \quad \yv=\dfrac{\wv}{t},
\end{align}
where $\yv=[y_0, \hdots, y_n]^{\T}$ and $\bv=[b_0, \hdots, b_n]^{\T}$, which results in the equivalent problem
\begin{align}
\text{(P3'($S$))} \hspace{0.7cm} \max\limits_{u,\bv,\yv} \quad & \sum\limits_{i=0}^n b_i f(\vv^i) \notag \\
\text{s.t.} \quad & u\leq 1/\alpha, \label{P3:u_alpha} \\
& u \cdot g_{S}(\yv/u)\leq 1, \label{P3:u_gS} \\
& \yv=\sum\limits_{i=0}^n b_i\vv^i, \label{P3:link_y_b} \\
& \sum\limits_{i=0}^n b_i-u=0, \label{P3:link_b_u} \\
& b_i \geq 0, \sp i=0,\hdots,n, \label{P3:dom_b} \\
& u > 0. \notag
\end{align}

Since $u\cdot g_{S}(\yv/u)$ is an affine function, (P3'($S$)) is a linear program, except for the domain constraint on $u$. However, \cite{Avr1988} showed that when a solution to (P2($S$)) exists, then the strict inequality can be replaced by $u\geq 0$, and we obtain the linear program
\begin{align}
\text{(P3($S$))} \hspace{0.7cm} \max\limits_{u,\bv,\yv} \quad & \sum\limits_{i=0}^n b_i f(\vv^i) \notag \\
\text{s.t.} \quad & \eqref{P3:u_alpha}-\eqref{P3:dom_b}, \notag \\
& u \geq 0. \label{P3:dom_u}
\end{align}
This formulation can now be solved very efficiently using any linear programming solver.

Finally, the upper bound for $S_l^k$, $l=1,2$, is now computed as $\min \{UB(S_l^k), UB(S^k)\}$.
Moreover, for each iteration $k\geq 0$, the upper bounding process also computes an upper bound $UB_k$ for the global optimal value $h(\wv^*)$ of the original problem~(P) based on the partition $\Q_k$:
\begin{align}
  UB_k=\max\limits_{S \in \Q_k} UB(S)\,.
\end{align}
By construction, the upper bound is monotonically decreasing in $k$, i.e., $UB_{k+1} \leq UB_k$, $k \geq 0$.

\subsubsection{Lower bounding process}

Denoting by $\wv^k$ the best solution of the problems (P1($S$)) encountered up to iteration $k$, the lower bound $LB_k$ for the global optimal value~$h(\wv^*)$ in iteration $k$ is given by $LB_k=h(\wv^k)$. The bounds are monotonically increasing in $k$: $LB_{k+1}\geq LB_k$, $k\geq 0$.

\subsubsection{Fathoming process}

Based on the lower and upper bounds produced by the algorithm, the fathoming process deletes all subsets $S \in \Q_{k-1}$ from $\Q_{k-1}$ that are guaranteed not to contain the global optimal solution. At the beginning of each iteration, i.e., all $S \in \Q_{k-1}$ are removed for which $(1-\rho)UB(S) \leq LB_{k-1}$.
If this results in $\Q_{k-1}$ being empty, then
\begin{align}
  h(\wv^{k-1}) \:\geq\: (1-\rho)UB_{k-1} \:\geq\: (1-\rho)\max_{\wv \in \W} h(\wv) \:=\: (1-\rho)h(\wv^*),
\end{align}
which means that $\wv^{k}$ is a $\rho$-globally optimal solution to problem (P). \cite{Ben2006} shows that when the number of iterations for the BB algorithm is infinite, it generates two sequences of points whose accumulation points are the global optimal solution $\wv^*$ for (P), and
\begin{align}
  \lim_{k \rightarrow \infty} LB_k=\lim_{k \rightarrow \infty} UB_k=h(\wv^*).
\end{align}
This result implies that whenever $\rho >0$, the BB algorithm is finite.
\medskip

\noindent
The complete BB algorithm is summarized below.
\begin{table}[h]
\begin{tabularx}{\textwidth}{X}
\toprule
\bf BB algorithm \\
\toprule
{\bf Input:} $\rho \in [0,1)$, $n$-simplex $\W$, functions $f(\cdot)$ and $g(\cdot)$. \\
{\bf Output:} $\rho$-globally optimal solution $\tilde{\wv}$. \\[1ex]
{\bf Initialization}
Set $S^0= \W$ and $\Q^0=\{S^0\}$. Calculate $UB_0=UB(S^0)$ and an optimal solution $(\wv^{0},t_{0})$ for P1($S^0$). Set $LB_0=h(\wv^0)$. \\
If $(1-\rho)UB_0\leq LB_0$, then stop; $\tilde{\wv}=\wv^0$ is $\rho$-globally optimal for (P). \\
\end{tabularx}
\begin{tabularx}{\textwidth}{X}
\vspace{-0.03cm}
{\bf Step $\boldsymbol{k}$} ($k=1,2, \hdots$)
\end{tabularx}
\begin{tabularx}{\textwidth}{l l X}
\vspace{-0.03cm}
\hspace{0.1cm} & $k.1$ & Delete each $n$-simplex $S \in \Q_{k-1}$ from $\Q_{k-1}$ for which $(1-\rho)UB(S) \leq LB_{k-1}$. \\
\hspace{0.1cm} & $k.2$ & If ${\Q}_{k-1}=\emptyset$, then stop: $\tilde{\wv}=\wv^{k-1}$ is $\rho$-globally optimal for problem (P). \\
\hspace{0.1cm} & $k.3$ &
Let $UB_k=\max\{UB(S) \mid S\in \Q_{k-1} \}$ and choose an $n$-simplex $S^k \in \Q_{k-1}$ such that $UB_k=UB(S^k)$.
Subdivide $S^k$ into two $n$-simplices $S_1^k,S_2^k$ via simplicial bisection.\\
\hspace{0.1cm} & $k.4$ & For $S=S_1^k, S_2^k$, find the optimal value $UB(S)$ and an optimal solution $(\wv^S,t_S)$ for P1($S$), and set $UB(S)=\min\{UB(S), UB(S^k)\}$. \\
\hspace{0.1cm} & $k.5$ & Set $LB_{k}=\max\{LB_{k-1}$, $h(\wv^{S_k^1}), h(\wv^{S_k^2})\}$ and let $\wv^{k}$ satisfy $LB_{k}=h(\wv^{k})$. Set $\Q_k={\Q}_{k-1} \setminus \{ S^k\} \cup\{S_1^k,S_2^k \}$ and $k=k+1$.\\
\bottomrule
\end{tabularx}
\end{table}

\noindent
\textbf{Remark.}
For the case with additional constraints, such as position limits, the feasible set is no longer given by the standard $n$-simplex. The extension of the algorithm to a more general case is however straightforward \citep[see][for details]{Ben2006}.

\subsubsection{Improving the upper bound}
\label{sec:BB:improved_upper_bound}
Preliminary computational tests showed that the BB algorithm spends the vast majority of the computing time calculating the upper bound $UB(S)$ over the $n$-simplex $S$. Moreover, while it often took only a few iterations to obtain a very good lower bound $LB_k$ on the optimal value $h(\wv^*)$, the upper bound was improving only very slowly. In order to achieve faster convergence for the BB algorithm, we present in the following two extensions of the algorithm presented in \cite{Ben2006} that lead to a much faster reduction of the global upper bound $UB_k$. In the first, the lower bound of the function $g$ in the denominator is improved by adding affine functions to the approximation. In the second, the upper bound of the function $f$ in the numerator is enhanced by using a generalization of the concave envelope. This generalization requires the introduction of binary variables, which means that the improved upper bound comes with the cost of having to solve a more difficult combinatorial optimization problem.

\begin{figure}
\begin{center}
\subfigure[]{
\resizebox*{.48\linewidth}{!}{\includegraphics{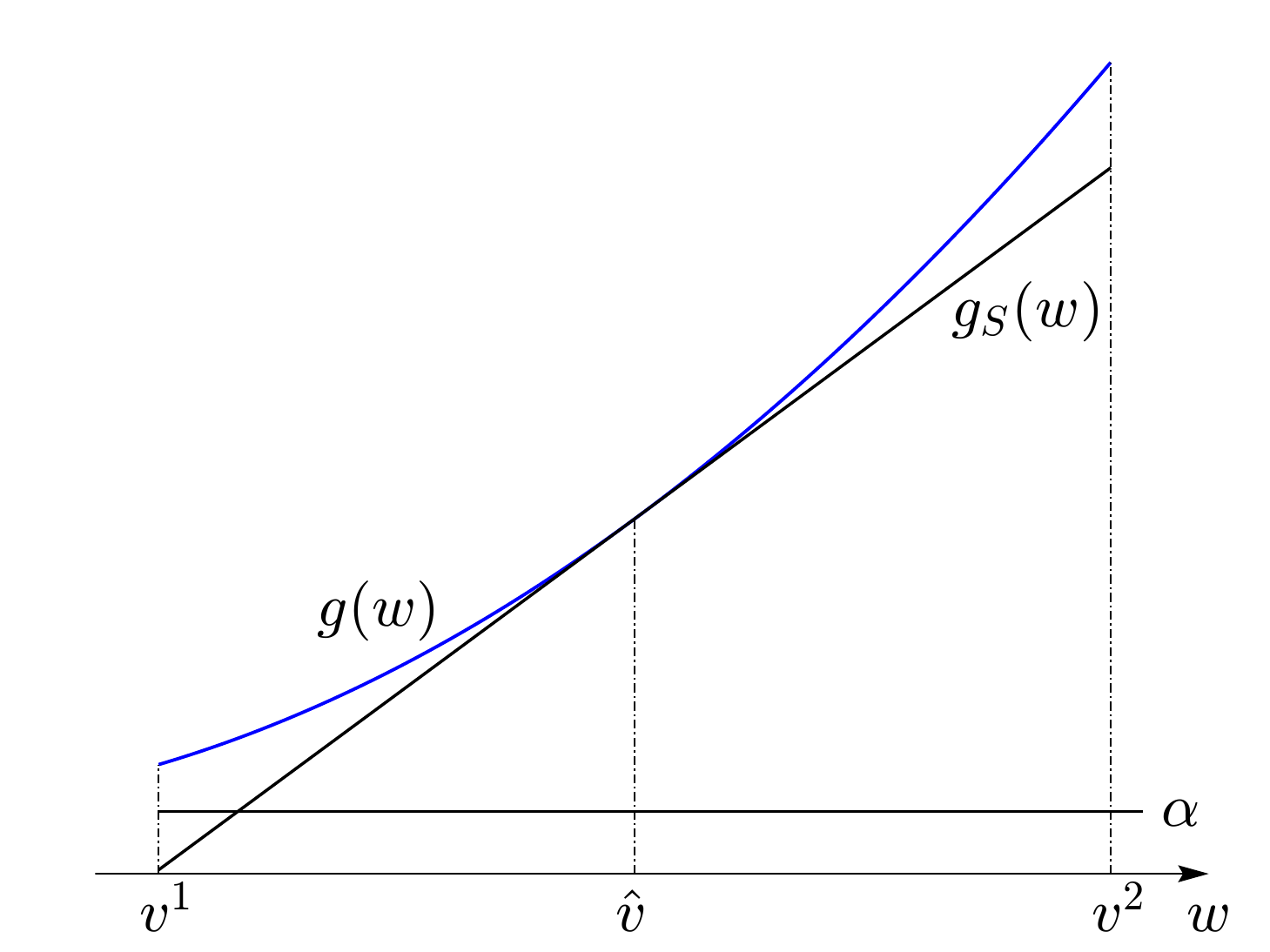}}}
\subfigure[]{
\resizebox*{.48\linewidth}{!}{\includegraphics{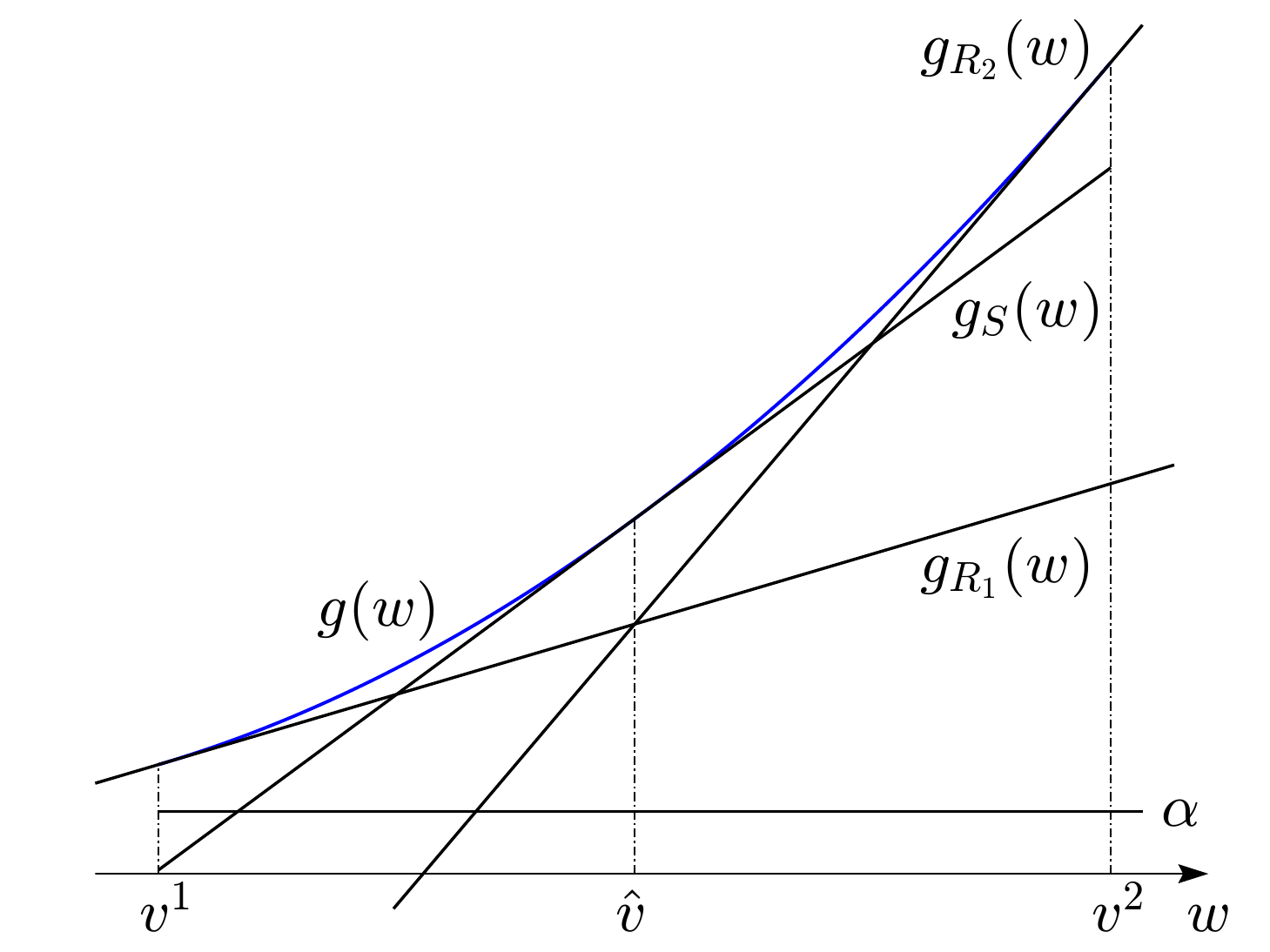}}}
\captionsetup{font=small}
\caption{Improvement of the lower bound for $g$ when $S$ is an  $1$-simplex: original lower bound (a) and improved lower bound for $p$=2 (b).}
\label{gBound}
\end{center}
\end{figure}

To tighten the lower bound for the function $g$, we extend the linearization technique in (\ref{zeq}) by adding first order Taylor expansions of $g$ around $p$ additional points $R_j$ in $S$. We then define
\begin{align}\label{ztilde}
  \tilde{z}(\wv)\:=\: \max(\alpha,g_S(\wv),g_{R_1}(\wv),\hdots,g_{R_p}(\wv)),
\end{align}
where $g_{R_j}(\wv)=g(\boldsymbol{R}_j)+\nw g(\boldsymbol{R}_j) (\wv-\boldsymbol{R}_j)^{\T}$, $j=1, \hdots,p$. The idea of the improvement is illustrated in Figure \ref{gBound} for the case when $S$ is a $1$-simplex and $p=2$.
For the general case of an $n$-simplex~$S$, the locations of the points $\{R_j\}_{j=1}^p$ are chosen so that they are evenly distributed in $S$, see Section~\ref{BBex} for more details. The resulting problem is then given as
\begin{align}
\text{(P4($S$))} \hspace{0.7cm} \max\limits_{u,\bv,\yv} \quad & \sum\limits_{i=0}^n b_i f(\vv^i) \notag \\
\text{s.t.} \quad & \eqref{P3:u_alpha}-\eqref{P3:dom_b}, \notag \\
& u \geq 0, \notag \\
& u\cdot g_{R_j}(\yv/u) \leq 1, \; j=0,\ldots,p. \label{P4:u_gSj}
\end{align}

Obviously, the accuracy of the approximation increases with $p$, at the expense of adding more linear constraints to the optimization problem.
\medskip

\begin{figure}
\begin{center}
\subfigure[]{
\resizebox*{.48\linewidth}{!}{\includegraphics{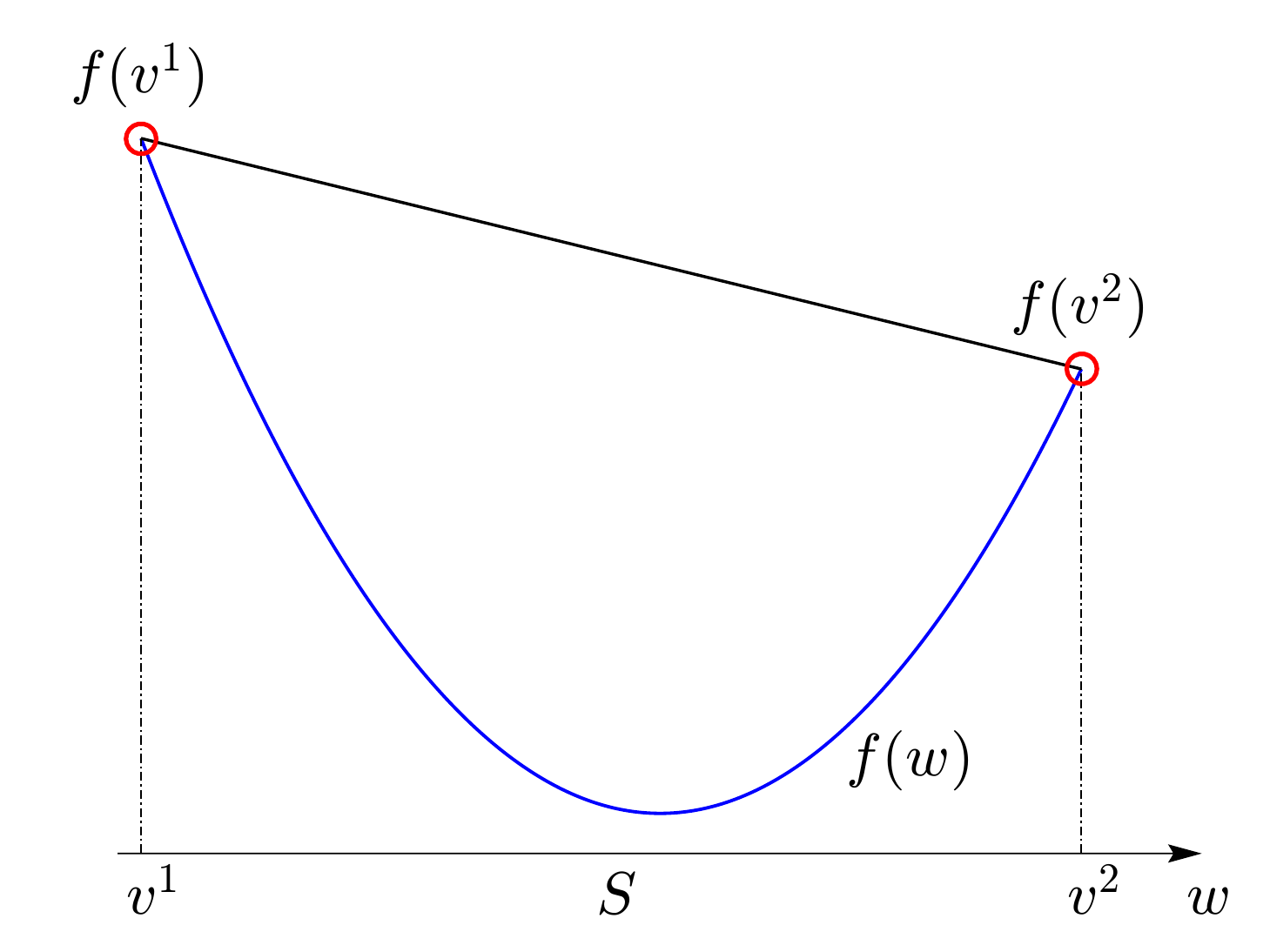}}}
\subfigure[]{
\resizebox*{.48\linewidth}{!}{\includegraphics{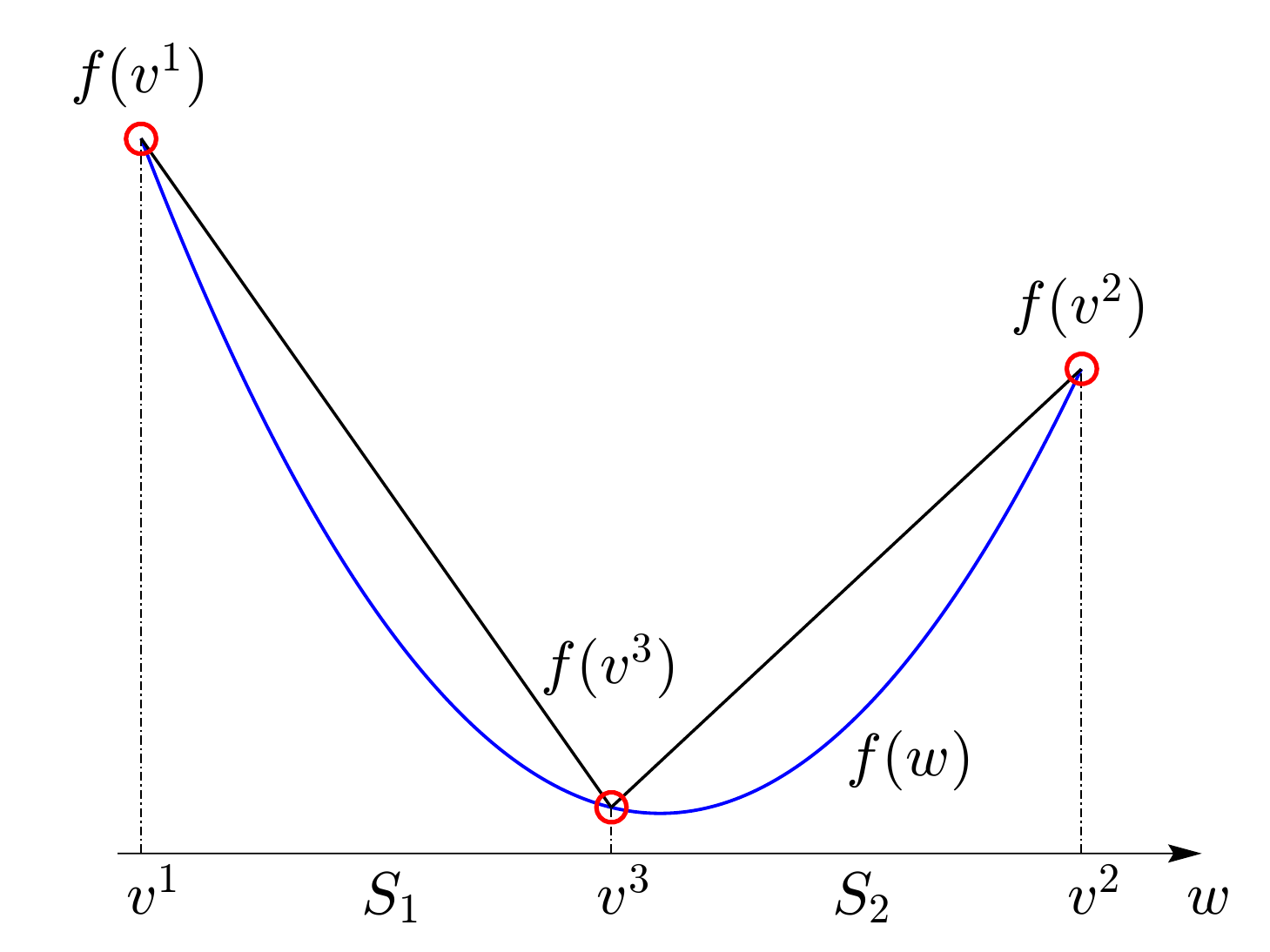}}\label{fBound}}
\captionsetup{font=small}
\caption{Improvement of the upper bound for $f$ when $S$ is an  $1$-simplex: original upper bound (a) and improved upper bound (b).}
\end{center}
\end{figure}

Next, we turn our attention to improving the accuracy of the approximation of the numerator of the objective function. We start by subdividing the $n$-simplex $S$ by radial subdivision according to Definition \ref{radsub}. Let the set of $n$-simplices created by the radial subdivision be given by $\mathcal{T}=\{ S_j\}_{j\in \J}$ and the corresponding set of all vertices by $\mathcal{V}(\mathcal{T})=\{\vv^i \}_{i\in \I}$.
The improved upper bound is then constructed by the combination of the concave envelopes over the $n$-simplices in $\Tp$.
The construction is more easily illustrated by the simplest possible example in one dimension given in Figure \ref{fBound}. For this example, the set of $n$-simplices and corresponding vertices are after the radial subdivision given by $\Tp=\{S_1,S_2\}$ and $\V(\Tp)=\{\vv^1,\vv^2,\vv^3 \}$, respectively. The generalized concave envelope over $S$ is constructed from the concave envelopes over $S_1$ and $S_2$.

When calculating the concave envelope, we have to introduce binary variables $q_j$, $j \in \J$, in order to keep track of which $n$-simplex in $\Tp$ is active. The function representing the generalized concave envelope over the $n$-simplex $S$ can now be formulated as
\begin{align}
& \sum\limits_{i\in \I } \lambda_i f(\vv^i) \notag \\
& \eqref{P2:sum_lambda}, \eqref{P2:dom_lambda}, \notag \\
& \sum\limits_{j\in \J}q_j=1, \label{ext_conc_env:sum_q} \\
& \lambda_i \leq \sum\limits_{j:\sp \vv^i \in S_j}q_j, \sp i \in \I, \label{ext_conc_env:link_lambda_q} \\
& q_j \in \{0,1\},\sp j \in \J. \label{ext_conc_env:dom_q}
\end{align}
Condition \eqref{ext_conc_env:link_lambda_q} ensures that only $\lambda_i$'s belonging to vertices of the $n$-simplex that is active, i.e. for which $q_j=1$, can be non-zero.
Using the improved approximation function, we obtain the optimization problem
\begin{equation*}
\begin{aligned}
\text{(P5($S$))} \hspace{0.7cm} &\max\limits_{t,\lv,\qv,\wv} && \dfrac{1}{t}\sum_{i\in \I} \lambda_i f(\vv^i) \\
& \text{s.t.} && \eqref{P1:t_alpha}, \eqref{P1:t_gS}, \eqref{P2:w}-\eqref{P2:dom_lambda}, \eqref{ext_conc_env:sum_q}-\eqref{ext_conc_env:dom_q}.
\end{aligned}
\end{equation*}
As before, we transform this problem into a mixed-integer linear program (MILP) via Charnes-Cooper transformation through the variable transformations in \eqref{varchange}.
The last set of constraints
is transformed into
\begin{align}
\label{P5:link_b_qu}
  b_i -\sum\limits_{j:\sp \vv^i \in S_j}q_ju\leq 0,\sp i \in \I,
\end{align}
in the new variables. The product of variables is linearized by introducing the continuous variables $z_j=q_j u$, $j \in \J$, and adding the following constraints for each $j \in \J$:
\begin{align}
\label{P5:model_z}
  & z_j \leq \dfrac{q_j}{\alpha}, \quad z_j \leq u, \quad
  z_j \geq u-(1-q_j)/\alpha, \quad\text{and}\quad z_j \geq 0.
\end{align}
If $q_j=0$, then the first and last constraint ensures that $z_j=0$, while the third only states that $z_j$ has to be greater than a negative number. If $q_j=1$, then the first constraint enforces $z_j \leq 1/\alpha$, and the second and third ensure that $z_j=u$.
Summing up, we obtain the following equivalent MILP, which we denote by (P6(S))
\begin{equation*}
\begin{aligned}
\text{(P6($S$))} \hspace{0.7cm} &\max\limits_{u,\bv,\zv,\qv,\yv} && \sum\limits_{i\in \I } b_i f(\vv^i)\\
& \text{s.t.} && \eqref{P3:u_alpha}-\eqref{P3:dom_b},  \eqref{ext_conc_env:sum_q}, \eqref{ext_conc_env:dom_q}, \eqref{P5:link_b_qu}, \eqref{P5:model_z}, \notag \\
& u \geq 0.
\end{aligned}
\end{equation*}
This problem can easily be enhanced by adding linear terms to the constraints in order to obtain a better approximation of the function $g$ in the denominator. The hope is that the improved upper bound in (P6($S$)) will lead to a sufficiently fast decrease of the global upper bound $UB_k$ in order to compensate for the increased computational time induced by solving an MILP instead of an LP for each instance of the upper bounding process.

\subsection{Numerical implementation}\label{BBex}

\begin{figure}
\begin{center}
\subfigure[]{
\includegraphics[width=.38\linewidth]{radialSubdivision}}
\hspace*{2cm}
\subfigure[]{
\includegraphics[width=.38\linewidth]{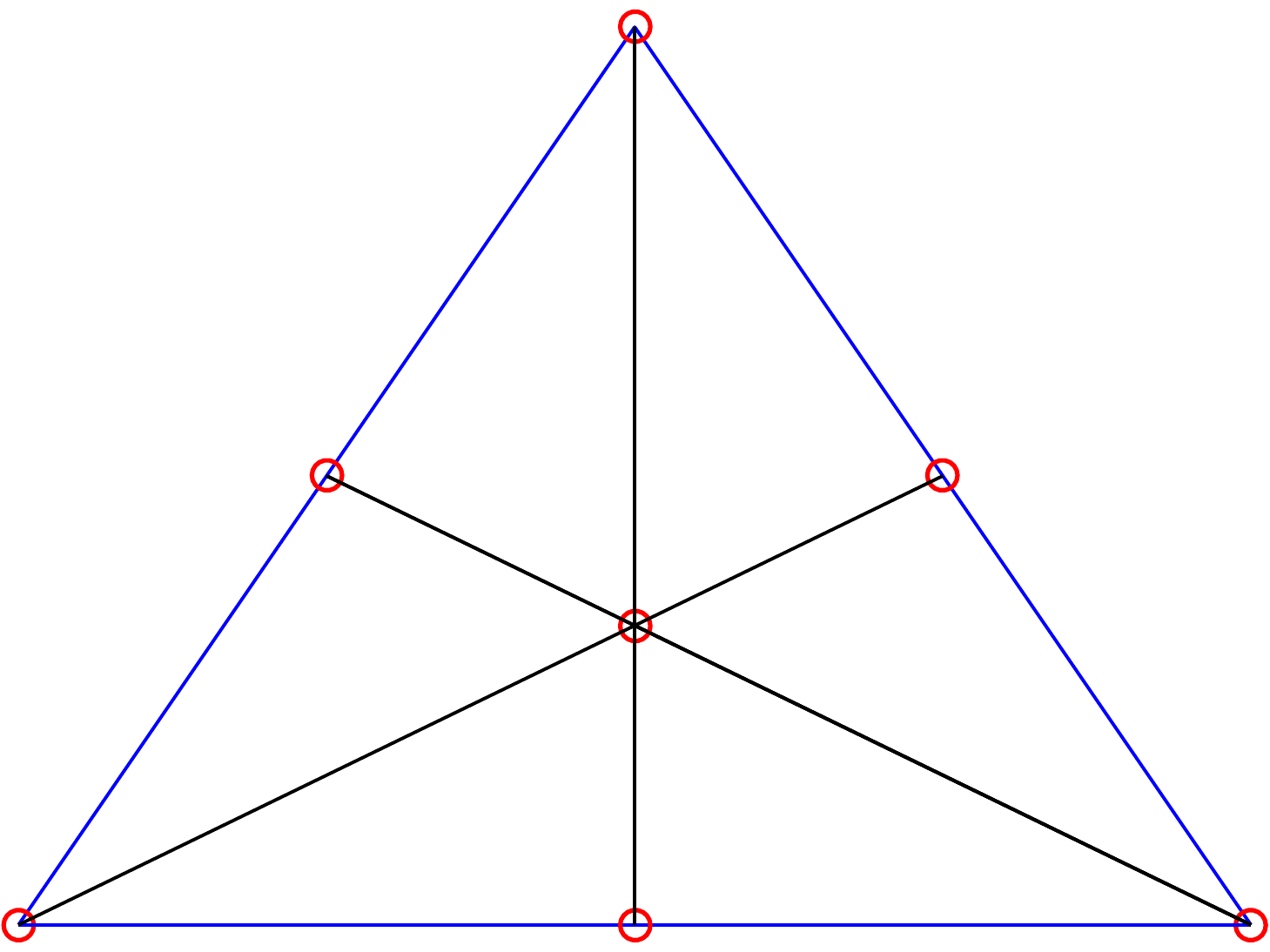}}
\captionsetup{font=small}
\caption{Examples of subdivision of a 2-simplex: radial subdivison (a) and barycentric subdivison (b).}
\label{barysubdivfig}
\end{center}
\end{figure}

In this section we will demonstrate the convergence properties of the BB algorithm when applied to the problem of minimizing the portfolio kurtosis for an increasing number of assets. As sample problem we assume that all assets have identical marginal distributions and that all correlations between different assets are assumed to be equal. This problem instance represents a non-convex problem with multiple local optima. When the subproblems for the BB algorithm are given by the MILP (P6($S$)), the description in Section \ref{BB} needs to be extended in order to produce an efficient algorithm. Numerical experiments show that radial subdivision does not improve the upper bound of $f$ sufficiently to produce an efficient algorithm. In order to further improve the upper bound of $f$, the $n$-simplex $S$ is instead subdivided with barycentric subdivision. Roughly speaking, the barycentric subdivision of an $n$-simplex $S$ is obtained by radial subdivision of all $k$-faces of dimension $1\leq k \leq n$ in decreasing order of dimension. It is also possible with partial barycentric subdivision of $S$ by restricting the radial subdivision to all $k$-faces of dimension $l \leq k \leq n$, with $l>1$, in decreasing order of dimension  \citep[see][for a detailed description of barycentric subdivision]{Ahm2018}. The partial barycentric subdivision of a 2-simplex with $l=2$ corresponds to radial subdivision as illustrated in Figure \ref{barysubdivfig} (a). The full barycentric subdivision of the 2-simplex is illustrated in Figure \ref{barysubdivfig} (b). Numerical experimentation reveals that full barycentric subdivision is required in order to produce a sufficiently improved upper bound of $f$ for the MILP formulation. Unfortunately, this means that $(n+1)!$ binary variables need to be introduced when solving the subproblems with the MILP formulation. The formulation of the optimization problem (P6($S$)) does not change when subdividing the $n$-simplex with barycentric subdivision instead of with radial subdivision. The only thing that changes is the set of $n$-simplices created by the subdivision and the corresponding set of vertices.

In the following we investigate the improvements obtained in terms of iteration count and runtime, when using the enhanced LP formulation and the MILP formulation for solving the subproblems of the algorithm. The BB algorithm was implemented in MATLAB. For all LP formulations of the subproblems we use the solver CPLEX in the implementation, whereas the subproblems arising from the MILP formulation are solved with the built-in solver {\it intlinprog} in MATLAB. For all comparisons we set the parameter $\rho$ to $10^{-3}$. When using the enhanced LP formulation (P4($S$)), a choice has to be made regarding how many extra constraints $p$ are added to the problem. The $p$ points defining the added constraints are distributed evenly over the subsimplex $S$ for which the subproblem is solved. Letting $n_c$ denote the number of added constraints per asset, one has that $p=(n+1)n_c$. We choose to distribute the $p$ points evenly between the vertices $\{\vv^i \}_{i=0}^n$ and the barycenter $\hat{\vv}$ of $S$. Thus, for $n_c=1$ the $p$ points are defined by the vertices. For $n_c\geq 2$, the $p$ points are defined by the vertices and the $(n+1)(n_c-1)$ points
\begin{align}
  \dfrac{j}{n_c}\vv^i+\left(1-\dfrac{j}{n_c}\right)\hat{\vv}, \sp i=0,\hdots,n; \sp j=1,\hdots, n_c-1.
\end{align}
We also experimented with distributing points evenly between the vertices but that did not bring any noticeable improvement in terms of iteration count. Naturally, adding more constraints in order to obtain a tighter lower bound should decrease the iteration count for the BB algorithm, at the cost of increasing the runtime for each of the subproblems that are solved. This trade-off is now investigated. In the following, we denote the enhanced LP formulation (P4($S$)) by LP2 and the LP formulation (P3($S$)) as used in \cite{Ben2006} by LP1.
\begin{figure}
\begin{center}
\subfigure[]{
\includegraphics[width=.48\linewidth]{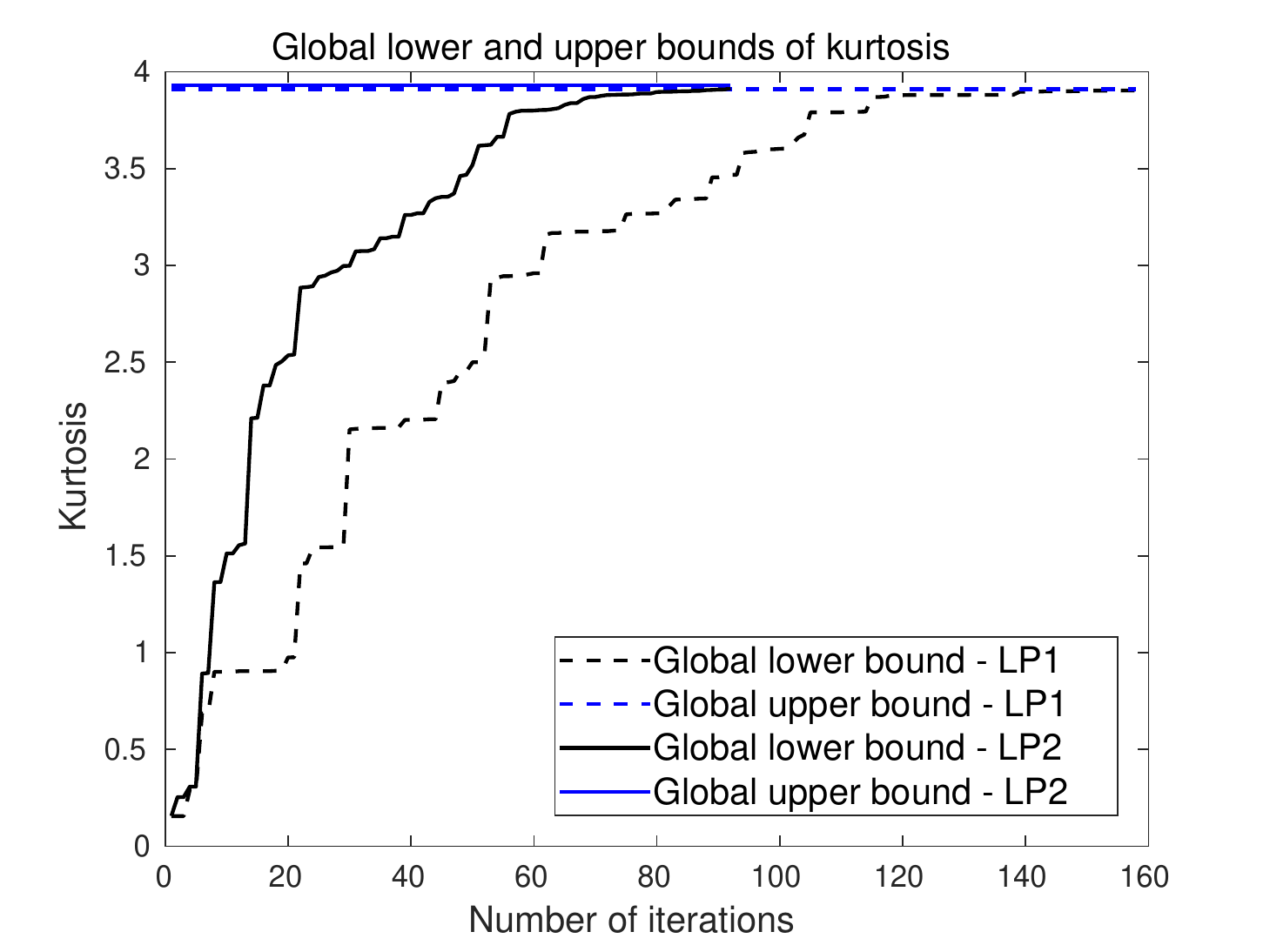}}
\subfigure[]{
\includegraphics[width=.48\linewidth]{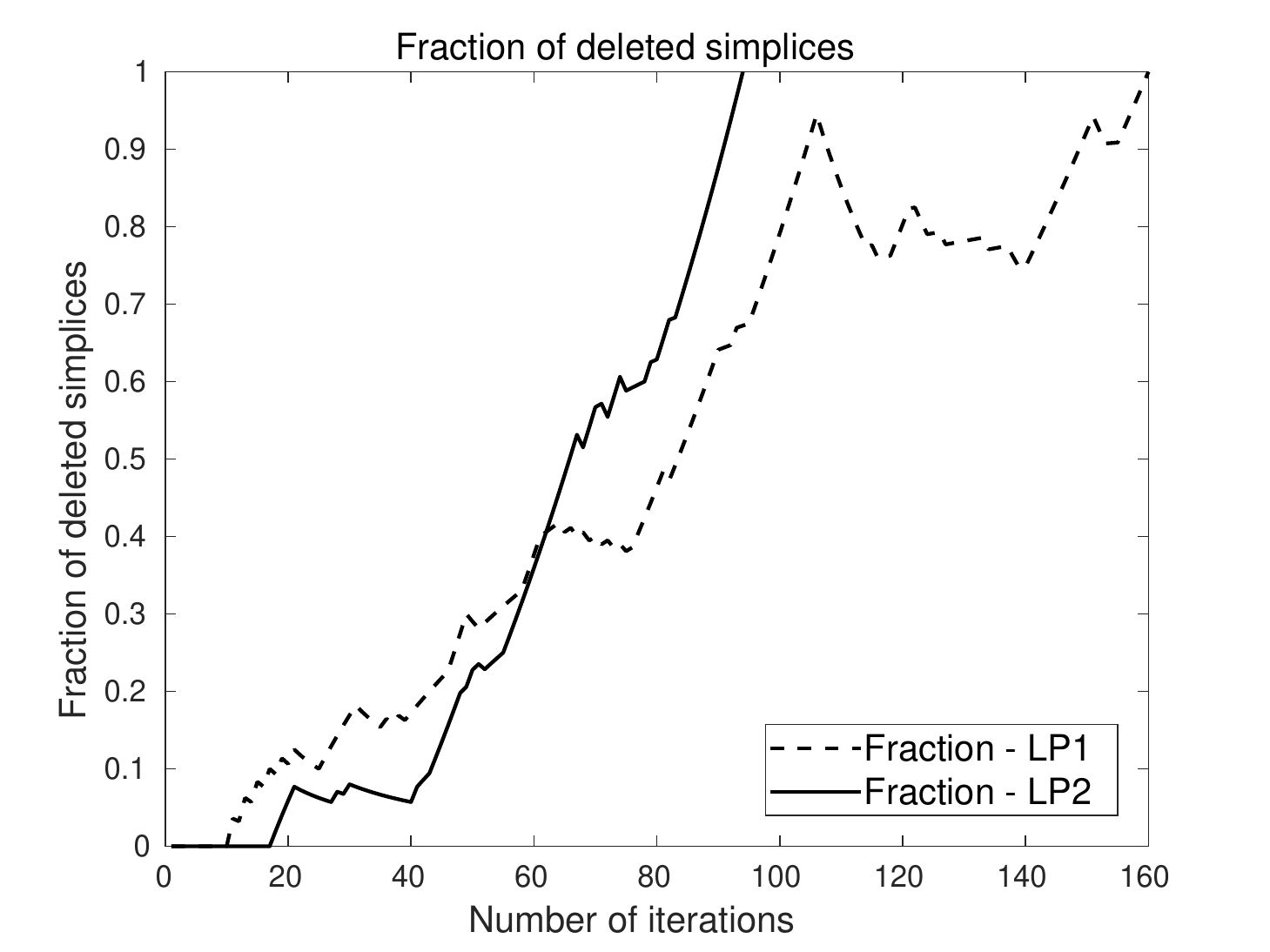}}
\captionsetup{font=small}
\caption{(a) Evolution of the global lower and upper bounds of the portfolio kurtosis for the original (LP1) and enhanced LP model (LP2) with $n_c=2$ for the three asset problem. (b) The fraction of deleted simplices for the original and enhanced LP model for the three asset problem.}
\label{BBiter3asset}
\end{center}
\end{figure}

Figure \ref{BBiter3asset} (a) displays the evolution of the global lower and upper bounds of the portfolio kurtosis for the iterations of the BB algorithm applied to the three asset problem. Note that these are the inverses of the global lower and upper bounds calculated by the BB algorithm, i.e., for $\kappa_p(\wv)=g(w)/f(w)$. Simulated return data is used to calculate the moment matrices in the objective function (\ref{optdef}). The sample moment matrices $\hat{\mathbf{M}}_2$ and $\hat{\mathbf{M}}_4$ are calculated from $10^7$ simulated asset returns with NIG-distributed margins and dependence structure given by the Gaussian copula with a homogeneous correlation matrix. The problem instance is defined by the homogeneous correlation $\rho=-0.2$ and marginal kurtosis $\kappa_m=6$ for all the assets. Appendix \ref{simsec} contains a description of the simulation procedure. Figure \ref{BBiter3asset} (b) shows the fraction of deleted simplices for the iterations of the algorithm for the three asset problem. Note that the fraction of deleted simplices decreases if the number of deleted simplices is less than the number of simplices that are added by the subdivision procedure. From the graphs it is visible that the enhanced LP formulation LP2 with $n_c=2$ converges faster to the global optimum in terms of number of iterations compared to the original LP formulation LP1. As can be seen in Table \ref{itertab}, the number of iterations decreases with the number of extra constraints $p$ added to the problem. However, as displayed in Table \ref{runtimetab} the decrease in the number of iterations is not significant enough in order to compensate for the increased runtime associated with the larger number of constraints. Thus, the enhanced LP formulation with $p=1$ has a lower runtime than the formulations with $p>1$. Compared to the original LP formulation LP1, the runtime of LP2 with $n_c=1$ is the same for the three asset problem.
\begin{figure}
\begin{center}
\subfigure[]{
\resizebox*{.48\linewidth}{!}{\includegraphics{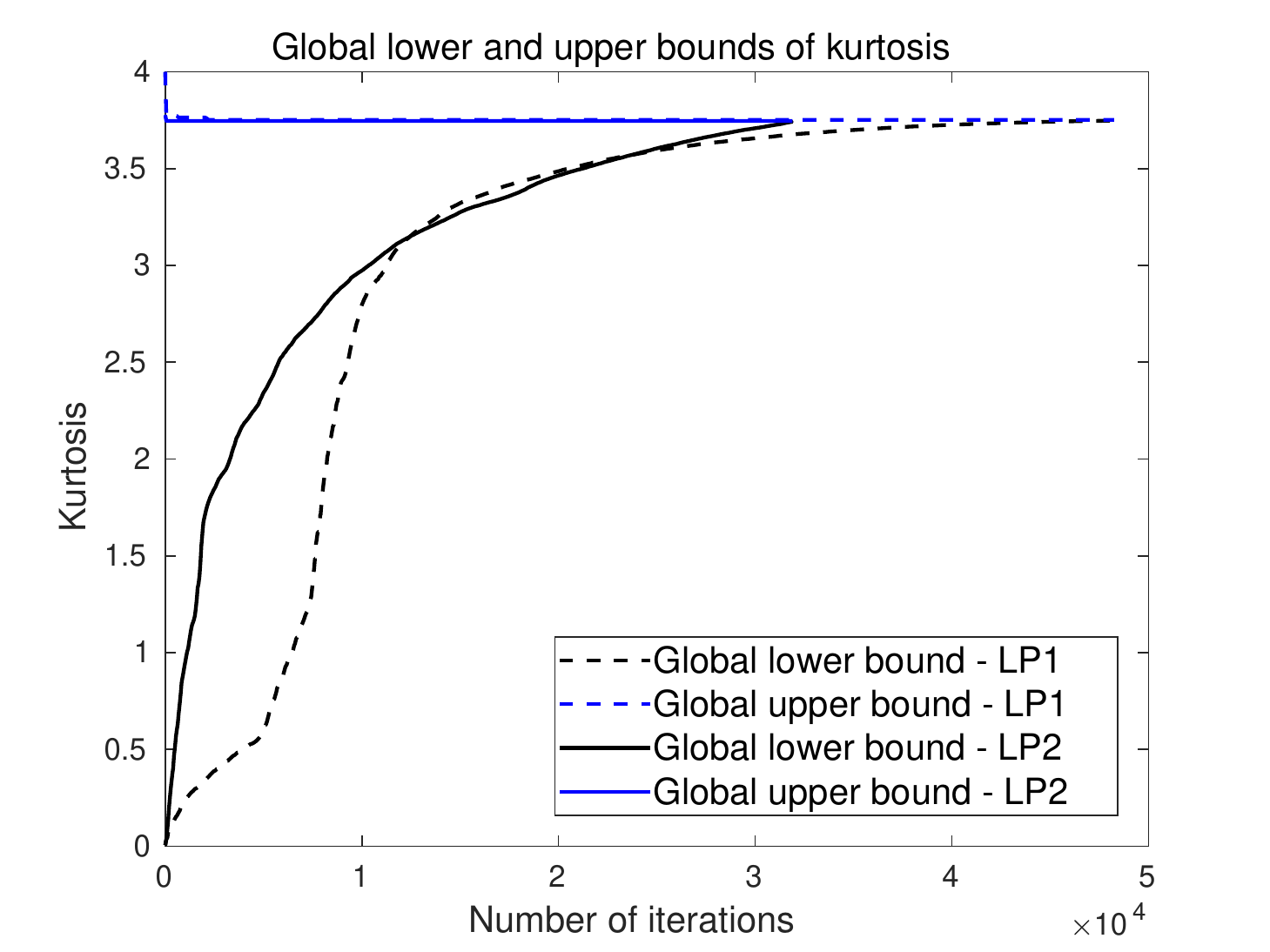}}}
\subfigure[]{
\resizebox*{.48\linewidth}{!}{\includegraphics{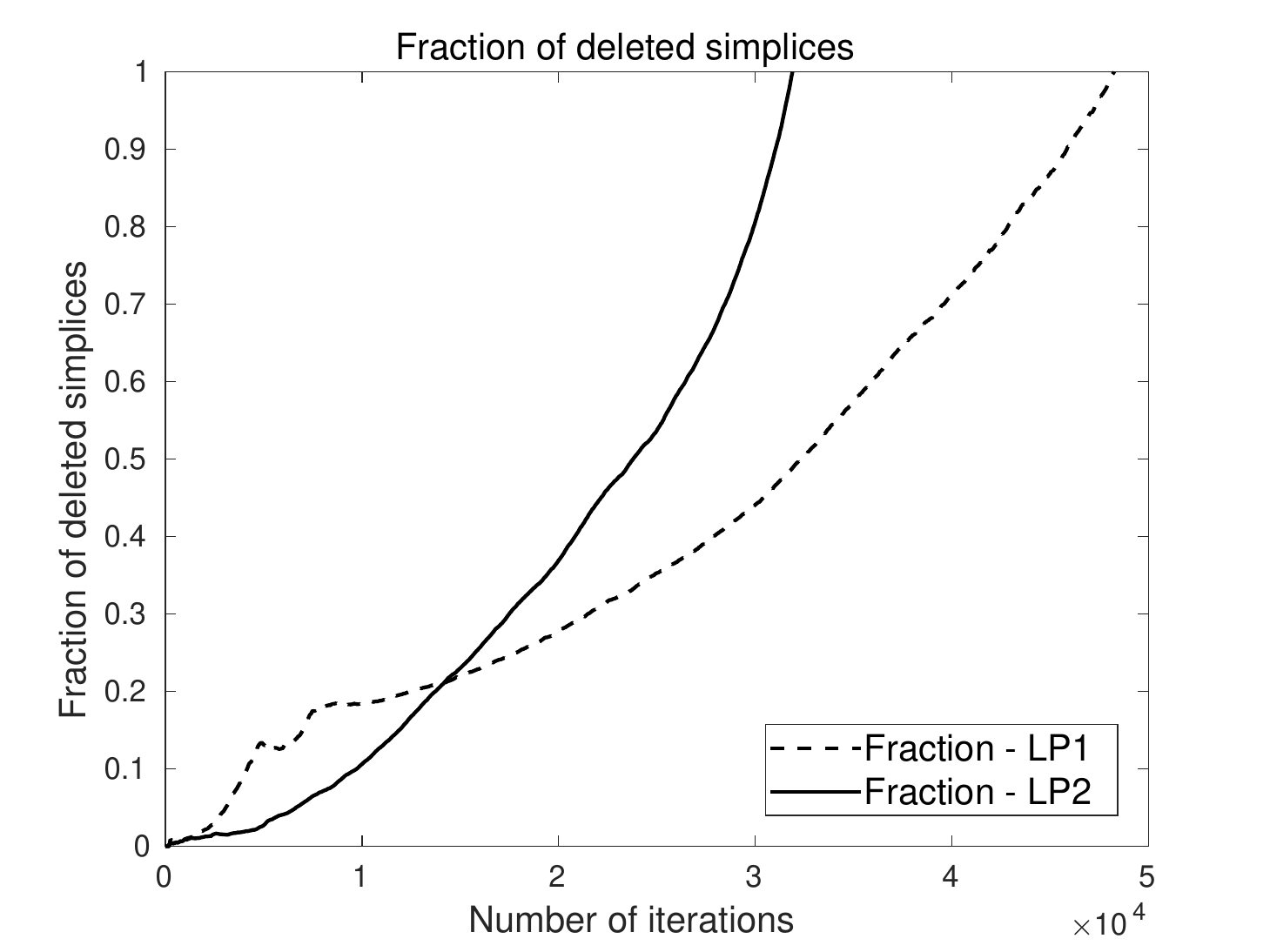}}}
\captionsetup{font=small}
\caption{(a) Evolution of the global lower and upper bounds of the portfolio kurtosis for the original (LP1) and enhanced LP model (LP2) with $n_c=2$ for the five asset problem. (b) The fraction of deleted simplices for the original and enhanced LP model for the five asset problem.}
\label{BBiter5asset}
\end{center}
\end{figure}

Figure \ref{BBiter5asset} (a) and (b) display the evolution of the global lower and upper bounds of the portfolio kurtosis and the fraction of deleted simplices for the iterations of the BB algorithm applied to the five asset problem. The solid and dotted lines represent the evolution of the bounds and the fraction of deleted simplices for LP2 with $n_c=2$ and LP1, respectively. The graphs reveal that there is a significant decrease in iteration count when using the enhanced LP formulation LP2 compared to LP1. As for the three asset case, the iteration count for LP2 decreases with the number of added constraints $p$. However, as can be seen in Table \ref{runtimetab} the decrease in iteration count does not compensate for the added computational cost and hence LP2 with $n_c=1$ has the lowest runtime. LP2 with $n_c=1$ also has a lower runtime than LP1 for the five asset case, 175 seconds compared to 193 seconds.
\begin{table}[h]
\begin{center}
  \begin{tabular}{ c  c  c c c c c}
    \hline \hline
    Subproblem method & LP1 & LP2 & LP2 & LP2 & LP2  & MILP  \\
                      &     & $n_c=1$& $n_c=2$& $n_c=4$& $n_c=8$ &      \\
    \hline
    3 assets          & 159 & 119 & 96 & 95 & 92 & 47 \\
    4 assets          & 1,551 & 1,762 & 1,358 & 1,254 & 1,264 & 1,068\\
    5 assets          & 48,254 & 35,943 & 33,374 & 30,824 & 29,041 & 28,179\\
    6 assets          &   -     &   620,000     &   -     &    -    &    -     &   -   \\
        \hline
  \end{tabular}
\end{center}
\captionsetup{font=small}
\caption{Number of iterations for the BB algorithm for different portfolio sizes and solution methods for the subproblems. The number of iterations for the enhanced LP model, LP2, is given for different numbers ($n_c$) of extra constraints per asset in the portfolio. In each case, the number of iterations is the median from five runs of the algorithm. The dashes in the table indicate that for the six asset problem, we have only produced results for the best performing algorithm in terms of runtime for the five asset problem.}
\label{itertab}
\end{table}

\begin{table}[h]
\begin{center}
  \begin{tabular}{ c  c  c c c c c}
    \hline \hline
    Subproblem method & LP1 & LP2 & LP2 & LP2 & LP2  & MILP  \\
                      &     & $n_c=1$& $n_c=2$& $n_c=4$& $n_c=8$ &      \\
    \hline
    3 assets          & 4 s & 4 s & 4 s & 5 s & 5 s & 6 s \\
    4 assets          & 8 s  & 10 s & 10 s & 12 s & 15 s & 61 s\\
    5 assets          & 193 s & 175 s & 198 s & 260 s & 425 s & 9,390 s\\
    6 assets          &   -     &  49,680 s      &    -    &   -     &    -     &   -   \\
        \hline
  \end{tabular}
\end{center}
\captionsetup{font=small}
\caption{Runtime for the BB algorithm for different portfolio sizes and solution methods for the subproblems. The runtime for the enhanced LP model, LP2, is given for different numbers ($n_c$) of extra constraints per asset in the portfolio. In each case, the runtime is the median from five runs of the algorithm. The dashes in the table indicate that for the six asset problem, we have only produced results for the best performing algorithm in terms of runtime for the five asset problem.}
\label{runtimetab}
\end{table}

\begin{figure}
\begin{center}
\subfigure[]{
\resizebox*{.48\linewidth}{!}{\includegraphics{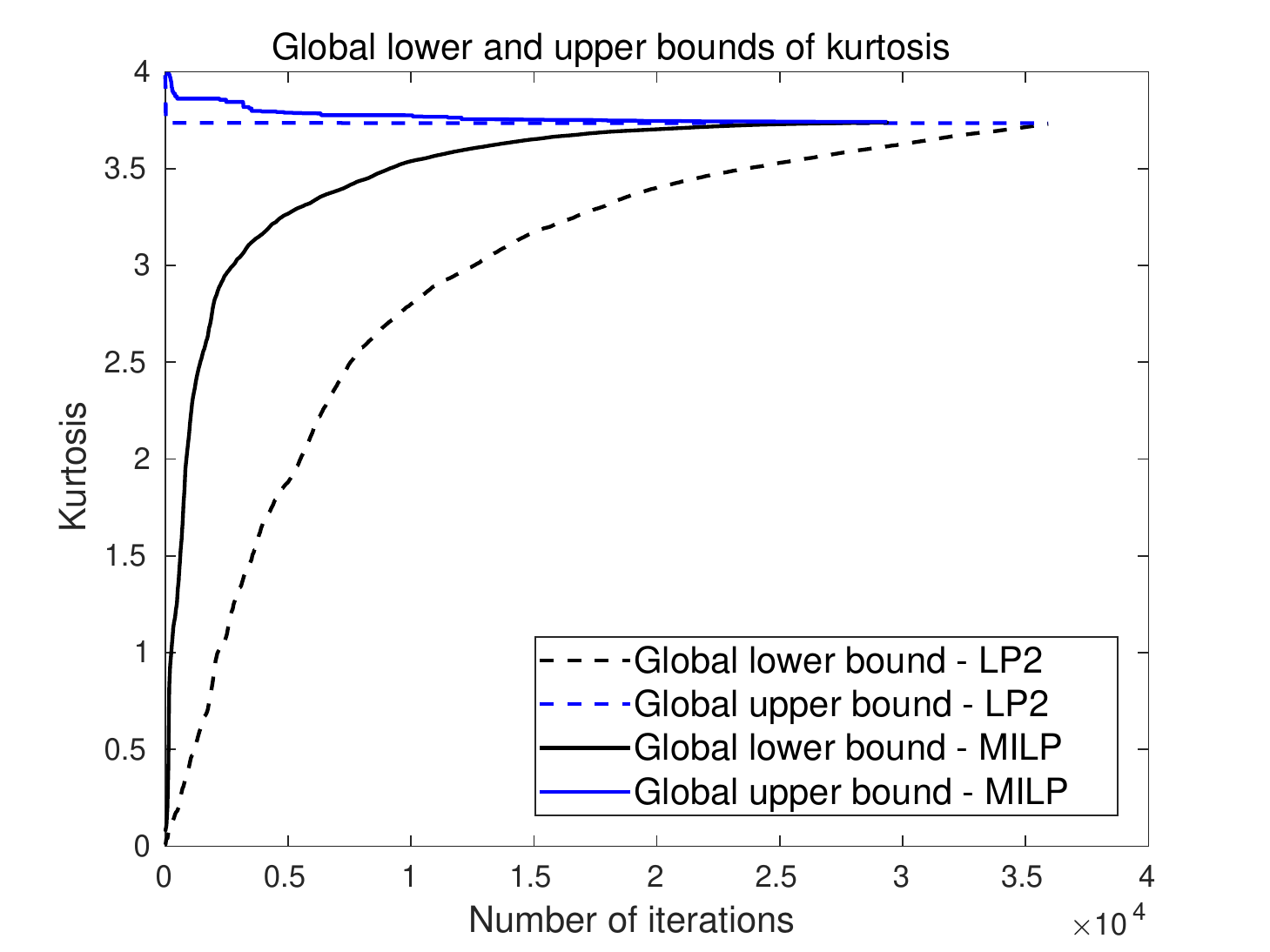}}}
\subfigure[]{
\resizebox*{.48\linewidth}{!}{\includegraphics{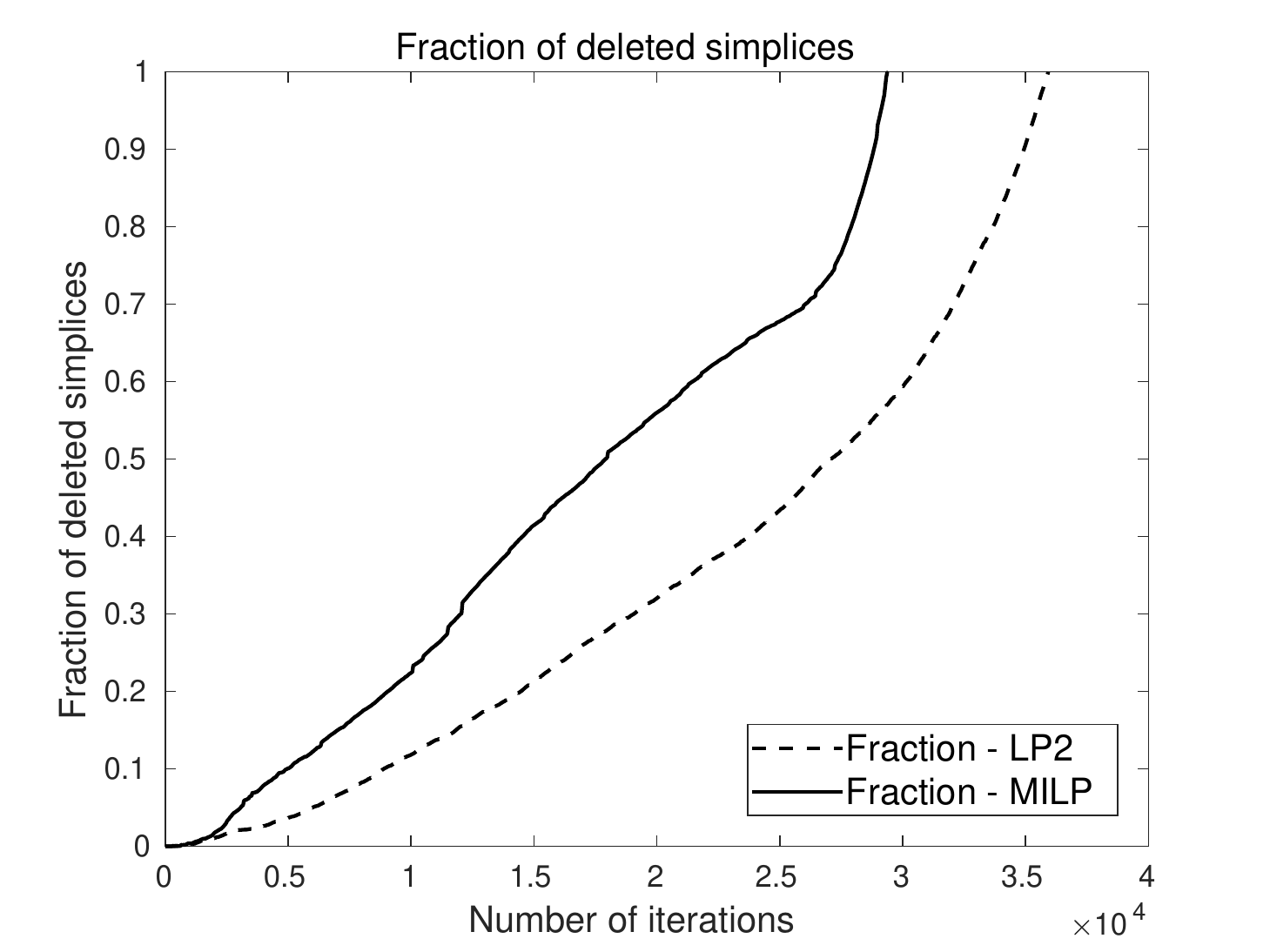}}}
\captionsetup{font=small}
\caption{(a) Evolution of the global lower and upper bounds of the portfolio kurtosis for the enhanced LP model with $n_c=1$ and the MILP model for the five asset problem. (b) The fraction of deleted simplices for the enhanced LP model and the MILP model for the five asset problem.}
\label{BBiter5assetMILP}
\end{center}
\end{figure}

We will now investigate the performance of the MILP formulation against the LP formulation with the lowest runtime for the five asset problem. Figure \ref{BBiter5assetMILP} (a) and (b) show the evolution of the global lower and upper bounds of the portfolio kurtosis and the fraction of deleted simplices for the two cases. The solid and dotted lines represent the evolution of the bounds and the fraction of deleted simplices for the MILP formulation and LP2 with $n_c=1$, respectively. From Figure \ref{BBiter5assetMILP}, one observes that the MILP formulation improves the global lower bound of the kurtosis, corresponding to the global upper bound for the BB algorithm, much faster than the LP formulation up to around iteration count 5,000. After that, the global lower bound of the MILP formulation improves slower than for the LP formulation. The overall iteration count is lower for the MILP formulation compared to LP2. However, the improvement in iteration count does not compensate for the increased computational cost associated with solving a MILP instead of an LP as can be seen in Table \ref{runtimetab}. The runtime for the MILP formulation can however likely be reduced by using a state-of-the art solver instead of the built-in solver in MATLAB.

For the six asset problem with homogeneous correlation $\rho=-0.18$, the number of iterations is 620,000 for the best performing solution method for the subproblems, LP2 with $n_c=1$. The corresponding runtime for the six asset problem is 49,680 seconds, illustrating the exponential growth in computational effort when the BB algorithm is applied to the portfolio kurtosis minimization problem. The BB algorithm can be enhanced by developing special purpose solvers for the subproblems. Furthermore, the algorithm can be parallelized in order to further reduce the runtime. Moreover, we could combine the MILP formulation and LP2, starting with the former to quickly raise the upper bound, and then switch to LP2 to save runtime. It is, however, unlikely that any of these will admit solving problems with significantly higher number of assets than six.

\section{Stochastic global optimization}
In Section \ref{BB} we developed a deterministic global optimization algorithm for minimizing the inverse of the introduced portfolio diversification measures. However, as is well known and illustrated by the numerical examples in Section \ref{BBex}, the BB algorithm suffers from the curse of dimensionality and converges too slowly for problems where the number of assets exceeds six. In this section we develop a stochastic optimization algorithm for global optimization of portfolio kurtosis. The BB algorithm has the desirable property that the objective function value at the obtained solution is guaranteed to be arbitrarily close to the global minimum. For the algorithm developed in this section it is not possible to determine if the solution is a global optimum. However, the algorithm is a special case of stochastic approximation with a rich and well developed theory for convergence analysis. Since the BB algorithm is limited to problems of moderate size, the algorithm developed in this section complements the BB algorithm in the sense that it allows for tackling problems of larger size.
\subsection{Stochastic algorithms for global optimization}\label{stochopt}
There is a huge literature on global optimization algorithms, so called metaheuristic methods, for which it is not possible to guarantee that the obtained solution is a global optimum. These methods iteratively search the feasible set for the global optimum and without prior knowledge there is always the possibility that the optimal point lies in an unexplored region when the algorithm stops. Important examples of metaheuristic methods are genetic algorithms \citep[][]{Hol1975}, simulated annealing \citep[][]{Kir1983} and tabu search \citep[][]{Glo1986}. The interested reader may consult \cite{Gen2010} for an overview of metaheuristic methods. Even though, for metaheuristic methods, it is not possible to guarantee that a global optimal point has been found, algorithms that are based on stochastic approximation have a solid theoretical foundation and in many cases non-asymptotic convergence results are available with explicit constants, see \cite{Dal2017}, \cite{Dur2017} and \cite{Dur2018}. This can be contrasted to many other popular metaheuristic methods where the theory is often incomplete or even nonexistent, see \cite{Spall2003}. A strong aspect of stochastic approximation is the rich convergence theory that has been developed over many years. It has been used to show convergence of many stochastic algorithms such as neural network backpropagation and simulated annealing. For a rigorous example where stochastic approximation methods are applied to problems in finance see \cite{Lar2012}.

In stochastic approximation one is concerned with finding at least one root $\tv^* \in \Tv^*\subseteq \R^d$ to $G(\tv)=0$,
based on noisy measurements of $G(\tv)$. Root finding via stochastic approximation was introduced in \cite{Rob1951} and important generalizations were made in \cite{Kie1952}. Consider the unconstrained minimization problem
\begin{align}\label{minL}
  \min\limits_{\tv} L(\tv),
\end{align}
where $L$ is a smooth function, which has multiple local minima. For the special case when $G(\tv)$ is given by $G(\tv)=\nabla_{\tv} L(\tv)$, the stochastic approximation algorithm is given by the following stochastic gradient descent (SGD) algorithm
\begin{align}\label{SGD}
  \tv_{k+1}=\tv_k-a_k H(\tv_k,\Xv_{k+1}),
\end{align}
where $\{\Xv_k\}_{k\in \mathbb{Z}}$ is a sequence of $\R^m$-valued i.i.d. data and $H(\tv_k,\Xv_{k+1})$ is an unbiased estimate of the gradient, i.e. $\nabla_{\tv} L(\tv)=\E(H(\tv,\Xv_{k+1})$. In (\ref{SGD}), $\{a_k\}$ can either be a decreasing positive sequence satisfying appropriate conditions or a fixed small positive value $a_k=\lambda>0$, for any $k\geq0$.

In many estimation problems, a full set of data is collected and $G$ (or $L$) is chosen by conditioning on the data. This conditioning removes the randomness from the problem and the estimation problem becomes deterministic. In the machine learning literature this is commonly referred to as the batch gradient descent algorithm, which is given by $\tv_{k+1}=\tv_k-a_k\bar{H}(\tv_k)$, where $\{\Xv_k\}_{k=1}^N$ is the collected data and
\begin{align}\label{batchgrad}
  \bar{H}(\tv)=\dfrac{1}{N}\sum\limits_{k=1}^NH(\tv,\Xv_k).
\end{align}
Since $L$ has multiple local minima, applying SGD to (\ref{minL}) may yield convergence to a local minimum of $L$. Under broad conditions, \cite{Kush1997} show that (\ref{SGD}) converges to one of the local minima of $L$ with probability 1. However, the iterates will often be trapped at a local optimum and will miss the global one. Nevertheless, SGD, or one of  its various extensions, is commonly used in machine learning for optimization of Deep Neural Networks, see \cite{Good2016}. When $L$ has a unique minimum, \cite{Chau2016} provide convergence results for the case with dependent data, discontinuous $L$, and fixed step size.

The idea behind simulated annealing is that by adding an additional noise term to the iterations one can avoid getting prematurely trapped in a local minimum of $L$. In \cite{Gel1991}, the following modified SGD algorithm is analyzed
\begin{align}
  \tv_{k+1}=\tv_k-a_k H(\tv_k,\Xv_{k+1})+b_k \ev_{k+1},
\end{align}
where $\{\ev_k \}$ is a sequence of standard $d$-dimensional independent Gaussian random variables, and $\{a_k\}$ and $\{b_k\}$ are decreasing sequences of positive numbers tending to zero. They show that under suitable assumptions, $\tv_k$ tends to the global minimizer as $k \rightarrow \infty$ in probability. In the machine learning literature, the closely related Stochastic Gradient Langevin Dynamics (SGLD) algorithm has attracted significant interest in the research community in recent years. The SGLD algorithm for global optimization can be formulated as
\begin{align}
  \tv_{k+1}=\tv_k-a_k H(\tv_k,\Xv_{k+1})+\sqrt{2a_k/\beta} \ev_{k+1},
\end{align}
where $\{\ev_{k}\}$ is a sequence of standard $d$-dimensional independent Gaussian variables and $\beta>0$ is a temperature parameter. The batch version of this algorithm, Gradient Langevin Dynamics (GLD), is correspondingly given by
\begin{align}
  \tv_{k+1}=\tv_k-a_k\bar{H}(\tv_k)+\sqrt{2a_k/\beta} \ev_{k+1}.
\end{align}
Assuming that the gradient $H$ is Lipschitz continuous and under further assumptions, \cite{Rag2017} provide a non-asymptotic analysis of SGLD and GLD applied to non-convex problems for the case when the step size $a_k$ is a positive constant. The analysis provides non-asymptotic guarantees for SGLD and GLD to find an approximate minimizer. The rate of convergence is further improved for both SGLD and GLD in the recent papers by \cite{Xu2018} and in \cite{Chau2019} even in the presence of dependent data streams.
\subsection{A Gradient Langevin Dynamics algorithm for minimization of kurtosis}
Motivated by the enormous progress in the aforementioned optimization algorithms, we develop a GLD algorithm for global minimization of portfolio kurtosis. Since portfolio kurtosis is the ratio of two convex functions, the batch gradient is not simply given by the average as in (\ref{batchgrad}). Given a sample of observed return data for a given asset universe, the sample covariance matrix and sample fourth co-moment matrix can be estimated. The batch version of the portfolio kurtosis is then given by
\begin{align}
  \bar{h}(\wv)=\dfrac{\bar{f}(\wv)}{\bar{g}(\wv)}=\dfrac{\wv^{\T}\hat{\mathbf{M}}_4(\wv \otimes \wv \otimes \wv)}{(\wv^{\T}\hat{\mathbf{M}}_2\wv)^2},
\end{align}
where $\hat{\mathbf{M}}_2$ and $\hat{\mathbf{M}}_4$ denote the sample covariance and fourth co-moment matrices, respectively. Given the complicated form of the approximate bias for sample kurtosis, see \cite{Bao2013}, global minimization of portfolio kurtosis is not easily adapted to the algorithms in Section \ref{stochopt} which utilize a stochastic unbiased estimate of the gradient. For this reason we only develop a GLD algorithm for the global minimization problem.

The algorithms in Section \ref{stochopt} are formulated for unconstrained optimization problems and hence need to be adapted to constrained minimization over the standard $n$-simplex. The GLD algorithm for the constrained problem is given by the following projected iterations
\begin{align}\label{kurtiter}
  \wv_{k+1}=\Pi_{\W}\left(\wv_k-\lambda\nw \bar{h}(\wv_k)+\sqrt{2\lambda/\beta}\ev_{k+1}\right),
\end{align}
where $\Pi_{\W}$ denotes the Euclidean projection onto the feasible set, $\lambda >0$ is the fixed step size and $\wv\in \R^{n+1}$. Euclidean projection of a point onto the standard $n$-simplex is a quadratic program which can be solved very efficiently. See \cite{Chen2011} for a fast and simple algorithm for computing the projection onto the standard $n$-simplex. The gradient of the batch version of portfolio kurtosis is given by
\begin{align}
  \nw \bar{h}(\wv)=\dfrac{\nw \bar{f} (\wv)}{\bar{g}(\wv)}-\dfrac{ \bar{f} (\wv)\nw \bar{g}(\wv)}{\left(\bar{g}(\wv)\right)^2},
\end{align}
where the explicit form of the gradient $\nw \bar{f}(\wv)$ is given in Appendix \ref{app1} and
\begin{align}\label{gnabla}
  \nw \bar{g}(\wv)=4\left(\wv^{\T}\hat{\mathbf{M}}_2\wv\right)\hat{\mathbf{M}}_2 \wv.
\end{align}

Most convergence results for SGLD and GLD are only applicable for algorithms without projection. A natural way to avoid the projection step in each iteration would be to extend the objective function with a convex function outside of the feasible set. Naturally, the extended objective function needs to be continuous on the boundary of the feasible set and have a continuous gradient on the boundary. However, in \cite{Taw2002} it is shown that a sufficient condition for the existence of a convex extension of a function outside of a convex feasible set, is the convexity of the function. Even if the requirement of convexity of the function to be extended is relaxed such that convexity is only required close to the boundary of the feasible set, this does not hold for portfolio kurtosis. It can easily be shown that portfolio kurtosis in general is a non-convex function on the boundary of the feasible set. Hence, it is not possible to find a convex extension of portfolio kurtosis outside of the feasible set $\W$.

When the objective function is convex, \cite{Bub2018} provide convergence results for the projected SGLD and GLD algorithms. In the case of a non-convex objective function, no convergence results for projected SGLD and GLD currently exist in the literature, to the best of our knowledge. It should however be mentioned that the analysis of SGLD and GLD algorithms is currently a very active research area which is gaining significant recognition amongst the Optimization and ML research community.

In order for the iterations (\ref{kurtiter}) to converge, the gradient $\nw \bar{h}(\wv)$ needs to be Lipschitz continuous on the domain given by the feasible set $\W$. The Hessian of $h$ is given by
\begin{align}\label{hessian}
  \nabla_{\wv}^2 \bar{h}(\wv)=&\dfrac{1}{\left(\bar{g}(\wv)\right)^3}\left(\left(\bar{g}(\wv)\right)^2\nabla_{\wv}^2\bar{f}(\wv)-\bar{g}(\wv)\left(\nw\bar{f}(\wv)\left(\nw \bar{g}(\wv) \right)^{\T}+\nw\bar{g}(\wv)\left(\nw \bar{f}(\wv) \right)^{\T} \right)+\right. \nl
  & \left. \bar{g}(\wv)\bar{f}(\wv)\nabla_{\wv}^2 \bar{g}(\wv)+2\bar{f}(\wv)\nw\bar{g}(\wv)\left(\nw \bar{g}(\wv) \right)^{\T} \right),
\end{align}
where $\nw \bar{f}(\wv)$ and $\nw^2 \bar{f}(\wv)$ are given in Appendix \ref{app1}, $\nw \bar{g}(\wv)$ is given in (\ref{gnabla}) and
\begin{align}
  \nw^2 \bar{g}(\wv)=12\left(\wv^{\T}\hat{\mathbf{M}}_2\wv\right)\hat{\mathbf{M}}_2.
\end{align}
In (\ref{hessian}), each component of the numerator is a polynomial of degree 10 and the denominator is a polynomial of degree 12. Since it is assumed that $\hat{\mathbf{M}}_2$ is positive definite, the minimum $c$ of $\bar{g}(\wv)$ is strictly positive over $\W$, and one has that $|\bar{g}(\wv)|\geq c>0$ for $\wv\in \W$. As each component of $\nabla_{\wv}^2 \bar{h}(\wv)$ is a continuous function its value is bounded on a closed compact set, and hence
\begin{align}\label{hessnorm}
  \| \nabla_{\wv}^2 \bar{h}(\wv)\|_2 \leq K, \text{ for all }\wv \in \W,
\end{align}
which, using the mean value theorem, implies
\begin{align}
  \|\nw \bar{h}(\uv)-\nw \bar{h}(\vv)  \|_2\leq K\|\uv-\vv\|_2, \text{ for all } \uv, \vv \in \W,
\end{align}
where the matrix norm in (\ref{hessnorm}) is defined as the Hilbert-Schmidt norm. Thus, the gradient of the portfolio kurtosis is Lipschitz continuous over the feasible set. In both \cite{Rag2017} and \cite{Xu2018} it is required that the objective function is dissipative in order for the convergence results to hold. The objective function $\bar{h}$ is dissipative on $\W$ if there exists constants $m>0$ and $b\geq 0$ such that
\begin{align}
  \wv^{\T} \nw \bar{h}(\wv) \geq m\|\wv\|_2^2-b, \text{ } \forall \wv \in \W.
\end{align}
Since the gradient of $\bar{h}(\wv)$ is a continuous function it is bounded over $\W$:
\begin{align}
  \| \nw \bar{h}(\wv) \|_2 \leq K_2, \text{ } \forall \wv \in \W.
\end{align}
Over the $n$-simplex $\W$, the Cauchy-Schwartz inequality implies
\begin{align}
  |\wv^{\T}\nw \bar{h}(\wv)|\leq \|\wv\|_2\|\nw \bar{h}(\wv) \|_2 \leq \|\nw \bar{h}(\wv) \|_2 \leq K_2,
\end{align}
and hence $\wv^{\T}\nw \bar{h}(\wv) \geq -K_2$. Furthermore, $a(\|\wv\|_2^2-1) \leq 0$, for $a>0$, implying
\begin{align}
  \wv^{\T}\nw \bar{h}(\wv) \geq a(\|\wv\|_2^2-1)-K_2=a\|\wv\|_2^2-(K_2+a)=a\|\wv\|_2^2-b,
\end{align}
and hence $\bar{h}(\wv)$ is dissipative over $\W$. This means that portfolio kurtosis satisfies the assumptions underlying the convergence results in the non-convex case for GLD and SGLD without projection. Even though we cannot rely on formal convergence results from the literature for GLD with projection, we will in the next section apply the projected GLD algorithm to some example problems with multiple local minima.
\subsection{Numerical illustration}\label{Sec4_3}
In this section we apply the projected GLD algorithm to an artificial problem of kurtosis minimization when all assets are assumed to have identical marginal distributions and where all correlations between different assets are assumed to be identical. This problem provides a test bed for testing if the projected GLD algorithm finds an optimal point which is close to the global optimum. The problem has several local optima which, for the non-zero weights, represent equally weighted portfolios with exposure to all or a subset of the assets. To see that this must be the case, consider two assets with non-zero weight at a local optimum. Since the assets are linearly dependent and have the same marginal kurtosis, the optimization will allocate equal weight to both assets when they are non-zero. Hence, all locally optimal portfolios assign equal weight to all weights that are non-zero at the respective local optimum.

\begin{figure}
\begin{center}
\subfigure[]{
\resizebox*{.48\linewidth}{!}{\includegraphics{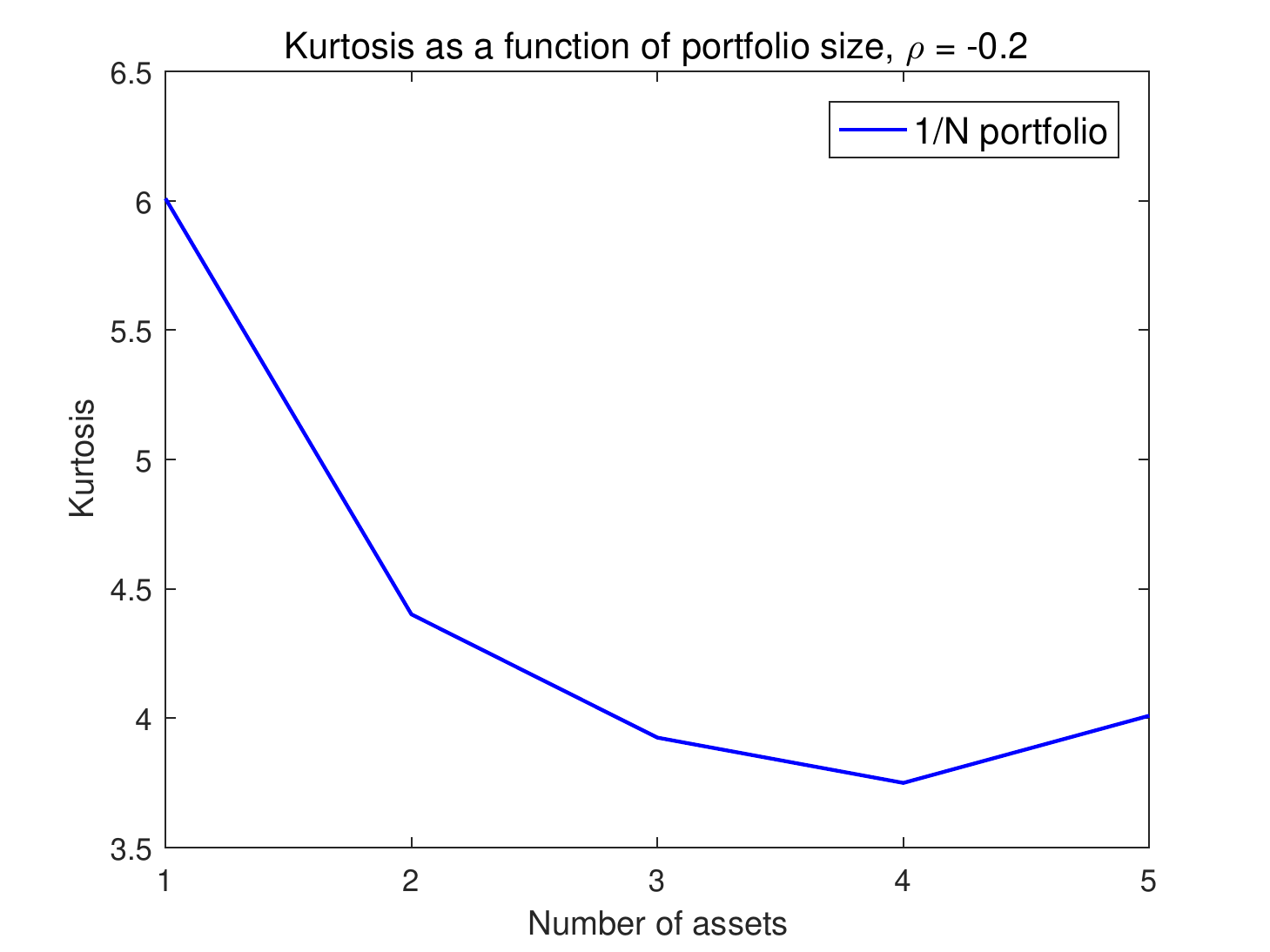}}}
\subfigure[]{
\resizebox*{.48\linewidth}{!}{\includegraphics{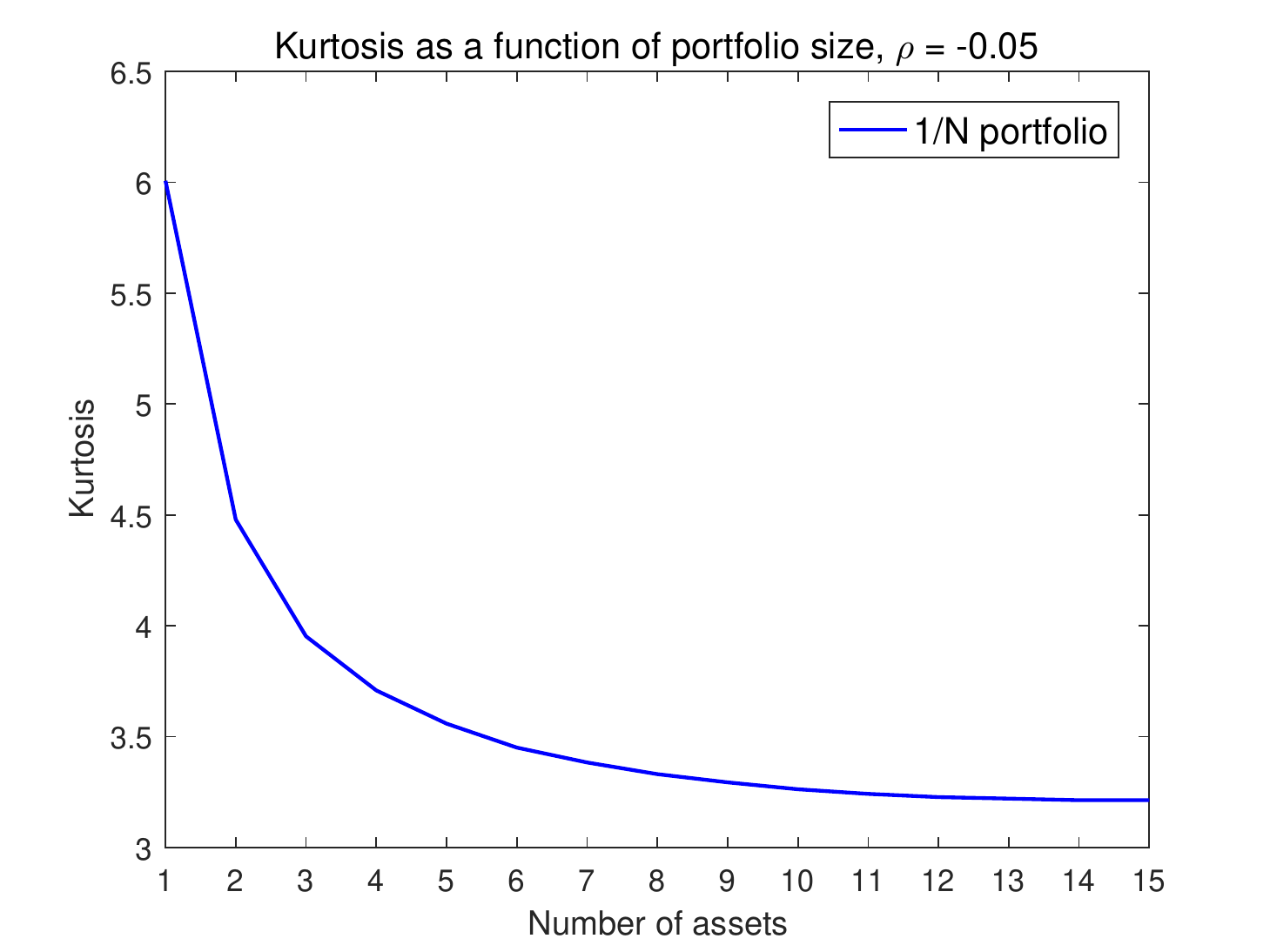}}}
\captionsetup{font=small}
\caption{Portfolio kurtosis for equally weighted portfolios as a function of the number of assets. The asset distributions are generated from a Gaussian copula with homogeneous correlation matrix and NIG-distributed margins with kurtosis $\kappa_m=6$. (a) Homogeneous correlation $\rho=-0.2$ (b) Homogeneous correlation $\rho=-0.05$.}
\label{Kurtfcnassets}
\end{center}
\end{figure}

The sample covariance matrix $\hat{\mathbf{M}}_2$ and the sample fourth co-moment matrix $\hat{\mathbf{M}}_4$ are calculated from $10^7$ simulated asset returns with NIG-distributed margins and dependence structure given by the Gaussian copula with a homogeneous correlation matrix. The simulation procedure is described in Appendix \ref{simsec}. Portfolio kurtosis for equally weighted portfolios for the cases with homogeneous correlation matrices with correlation $\rho=-0.2$ and $\rho=-0.05$, respectively, and marginal kurtosis $\kappa_m=6$, are displayed in Figure \ref{Kurtfcnassets}. Note that for the described experimental setup, assuming no estimation error, the portfolio kurtosis for an equally weighted portfolio with $n$ assets is equal to the portfolio kurtosis for an $n+1$ asset portfolio with equal weight in $n$ assets and zero weight in the remaining asset. To see this, consider the definition of portfolio kurtosis
\begin{align}
  \kappa_p(\wv)=\dfrac{\E\left(\wv^{\T} \rv \right)^4}{\left(\E\left(\wv^{\T} \rv \right)^2\right)^2}.
\end{align}
From the definition it is apparent that setting one of the weights to zero and the remaining weights to $1/n$ for the kurtosis in the $n+1$ asset case is identical to the kurtosis for the equally weighted portfolio in the $n$ asset case. Inspecting the graph in Figure \ref{Kurtfcnassets} (a), one observes that the global optima for the five asset case are located at points with equal weights in four of the assets and zero weight in the remaining asset. Thus, assuming no estimation error, there are five global optima for the five asset problem. With $10^7$ simulated sets of asset returns, the estimation error is small but nevertheless not zero and hence one of the points represents the unique global optimum with simulated data. Figure \ref{Kurtfcnassets} (b) displays the portfolio kurtosis for equally weighted portfolios with up to 15 assets for the case when the homogeneous correlation is -0.05. Even though not distinguishable from the graph, the portfolio kurtosis for the equally weighted portfolio with 14 assets is slightly lower than the equally weighted portfolio with 15 assets. Thus, for the 15 asset problem the global optimum for the kurtosis minimization problem is given by assigning equal weight to 14 of the assets and zero weight to the remaining asset.

The projected GLD algorithm is applied to the problem of minimizing portfolio kurtosis for the experimental setup described above, with five and 15 assets, respectively. We implement a multistart version of the algorithm, where the iterations (\ref{kurtiter}) are started from points $\wv_0\in \W$ uniformly sampled over the feasible set. In order to generate starting points that are evenly distributed over the $n$-simplex defining the feasible set, the method described in \cite{Shaw2010} is used. For each generated path of the projected GLD iterations, the point with the smallest recorded objective function value is stored. The output from the algorithm $\tilde{\wv}$ is the point with the smallest overall recorded objective function value. Finally, the optimal solution is taken to be
\begin{align}
  \wv^{\text{GLD}}=\argmin \{\kappa_p(\tilde{\wv}),\kappa_p(\wv^{\text{Loc}})\},
\end{align}
where $\wv^{\text{Loc}}$ denotes the solution from a local solver started at $\tilde{\wv}$. The complete multistart projected GLD algorithm is summarized below.
\begin{table}[h]
\begin{tabularx}{\textwidth}{X}
\toprule
\bf Multistart projected GLD algorithm \\
\toprule
{\bf Input:} $\lambda$, $\beta$, $n_{sim}$, $n_{iter}$, $\hat{\mathbf{M}}_2$, $\hat{\mathbf{M}}_4$. \\
\end{tabularx}
\vspace{-0.08cm}
\begin{tabularx}{\textwidth}{X}
{\bf for} $i=1,2,\hdots,n_{sim}$ {\bf do}
\end{tabularx}
\vspace{-0.01cm}
\begin{tabularx}{\textwidth}{X}
\hspace{0.5cm} Generate $\wv_0\in \R^{n+1}$ uniformly on $\W$; \\
\hspace{0.5cm} {\bf for} $k=0,1,\hdots,n_{iter}$ {\bf do} \\
\end{tabularx}
\begin{tabularx}{\textwidth}{X}
\hspace{1cm} Generate $\ev_{k+1}\sim N(\boldsymbol{0},\mathbf{I})$; \\
\hspace{1cm} $\wv_{k+1}=\Pi_{\W}\left(\wv_k-\lambda\nw \bar{h}(\wv_k)+\sqrt{2\lambda/\beta}\ev_{k+1}\right)$;\\
\end{tabularx}
\vspace{-0.01cm}
\begin{tabularx}{\textwidth}{X}
\hspace{0.5cm} {\bf end} \\
\hspace{0.5cm} $\wv_i^s=\argmin\{\kappa_p(\wv_0), \kappa_p(\wv_1), \hdots , \kappa_p(\wv_{n_{iter}}) \}$; \\
\end{tabularx}
\vspace{-0.04cm}
\begin{tabularx}{\textwidth}{X}
{\bf end}\\
{\bf Output:} $\tilde{\wv}=\argmin\{\kappa_p(\wv_0^s), \kappa_p(\wv_1^s), \hdots , \kappa_p(\wv_{n_{sim}}^s) \}$;\\
\bottomrule
\end{tabularx}
\end{table}
The fixed step size $\lambda$ is chosen to be 0.01 for both the five and 15 asset problems. The temperature parameter $\beta$ is chosen large enough so that the iterations from the projected GLD algorithm can jump between different local optima. Based on initial experimentation, the following formula for the temperature was chosen
\begin{align}
  \beta=\dfrac{2\lambda (n+1)^2}{c^2},
\end{align}
where $n+1$ is the number of assets and $c$ was chosen to be 0.06 for the five asset case and 0.1 for the 15 asset case. For the implementation, the number of paths $n_{sim}$ was $10^5$ and the number of iterations for each path $n_{iter}$ was $10^4$, implying that $10^9$ points in the search space were visited by the algorithm. Given the multistart implementation, the algorithm is very easy to parallelize. The algorithm was parallelized and implemented on a multi-core processor with 24 cores. For the five asset case, the multistart projected GLD algorithm finds a solution with equal weight in four assets and zero weight in one asset. By running the BB algorithm on the same problem it was confirmed that the GLD algorithm finds the global optimum. The runtime for the parallelized algorithm with 24 cores was 2,476 seconds for the five asset case and 8,322 seconds with 15 assets.
\begin{figure}
\begin{center}
\subfigure[]{
\resizebox*{.48\linewidth}{!}{\includegraphics{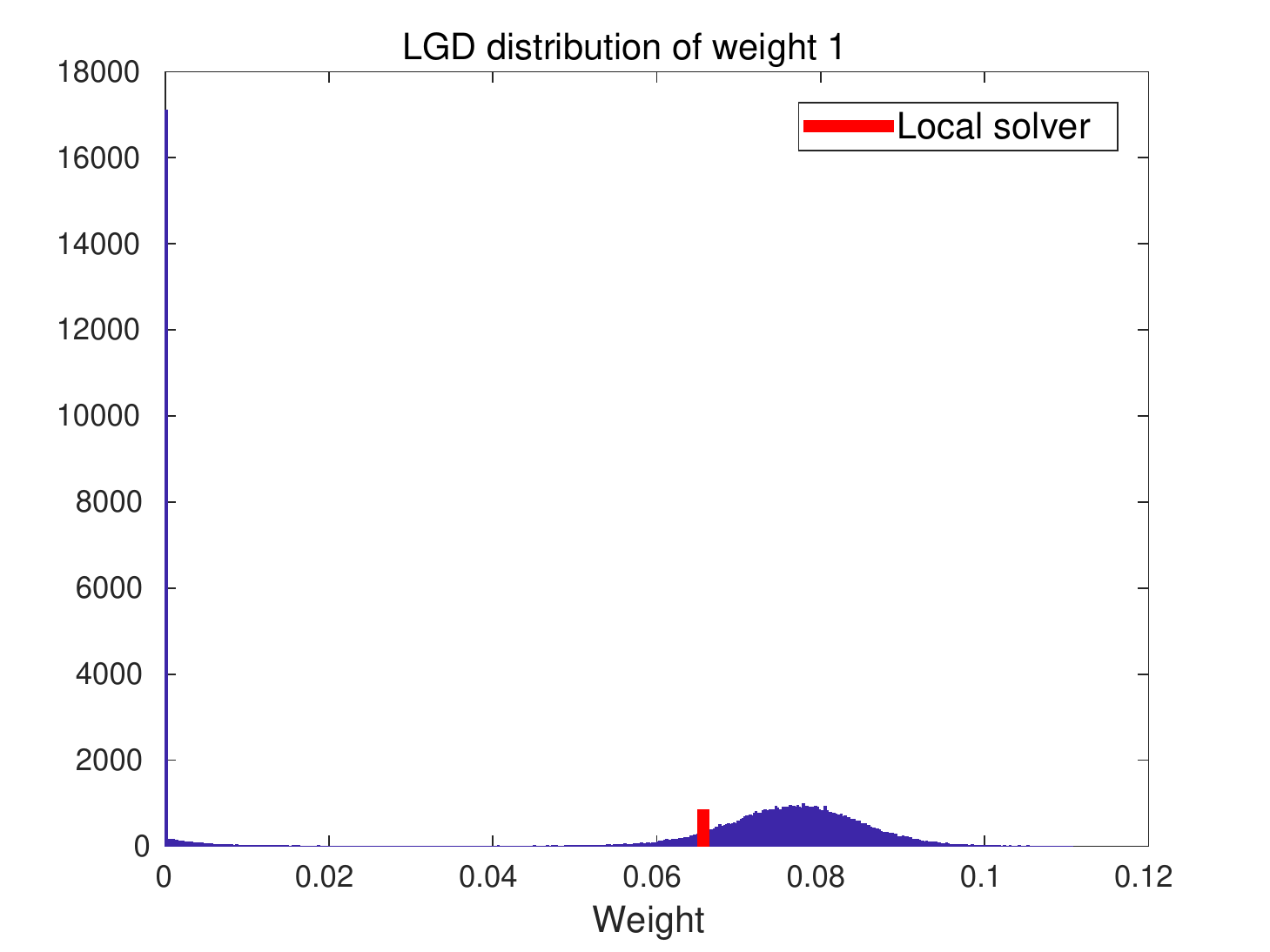}}}
\subfigure[]{
\resizebox*{.48\linewidth}{!}{\includegraphics{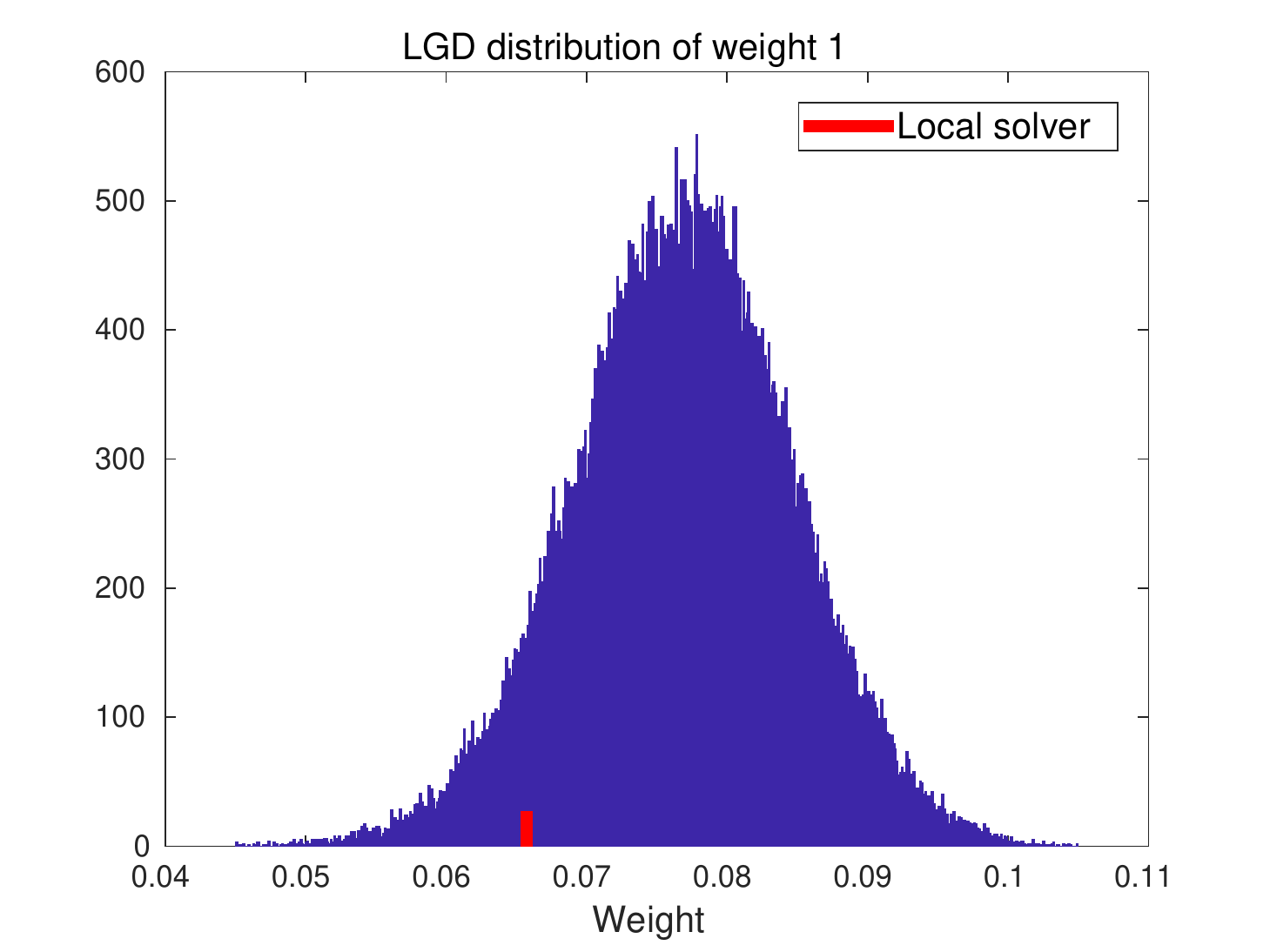}}}
\captionsetup{font=small}
\caption{The distribution of the final iterate for the weight of asset 1 in the 15 asset case: the full distribution (a) and the zoomed in distribution (b).}
\label{weightDist}
\end{center}
\end{figure}

\begin{figure}
\begin{center}
\subfigure[]{
\resizebox*{.48\linewidth}{!}{\includegraphics{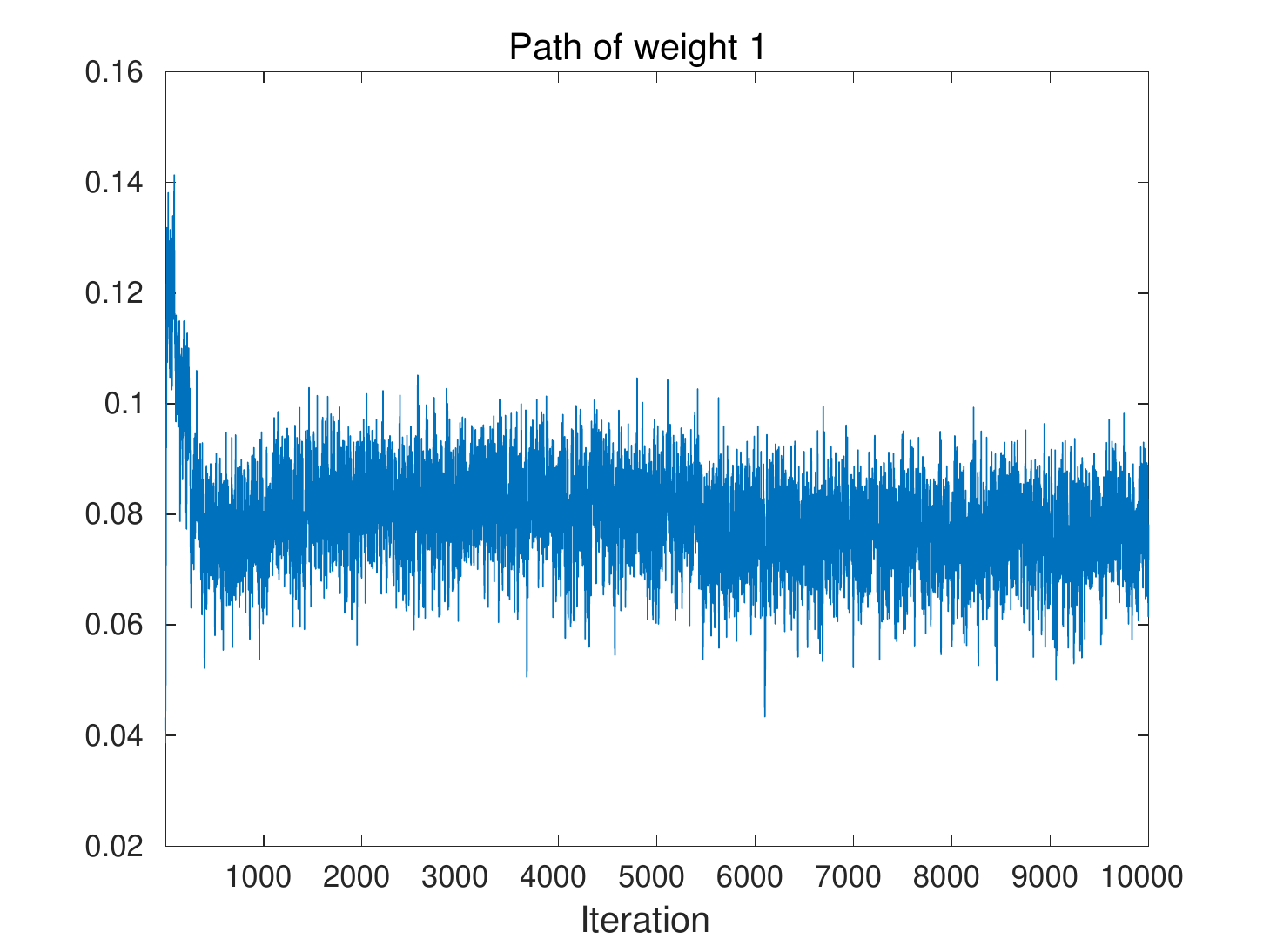}}}
\subfigure[]{
\resizebox*{.48\linewidth}{!}{\includegraphics{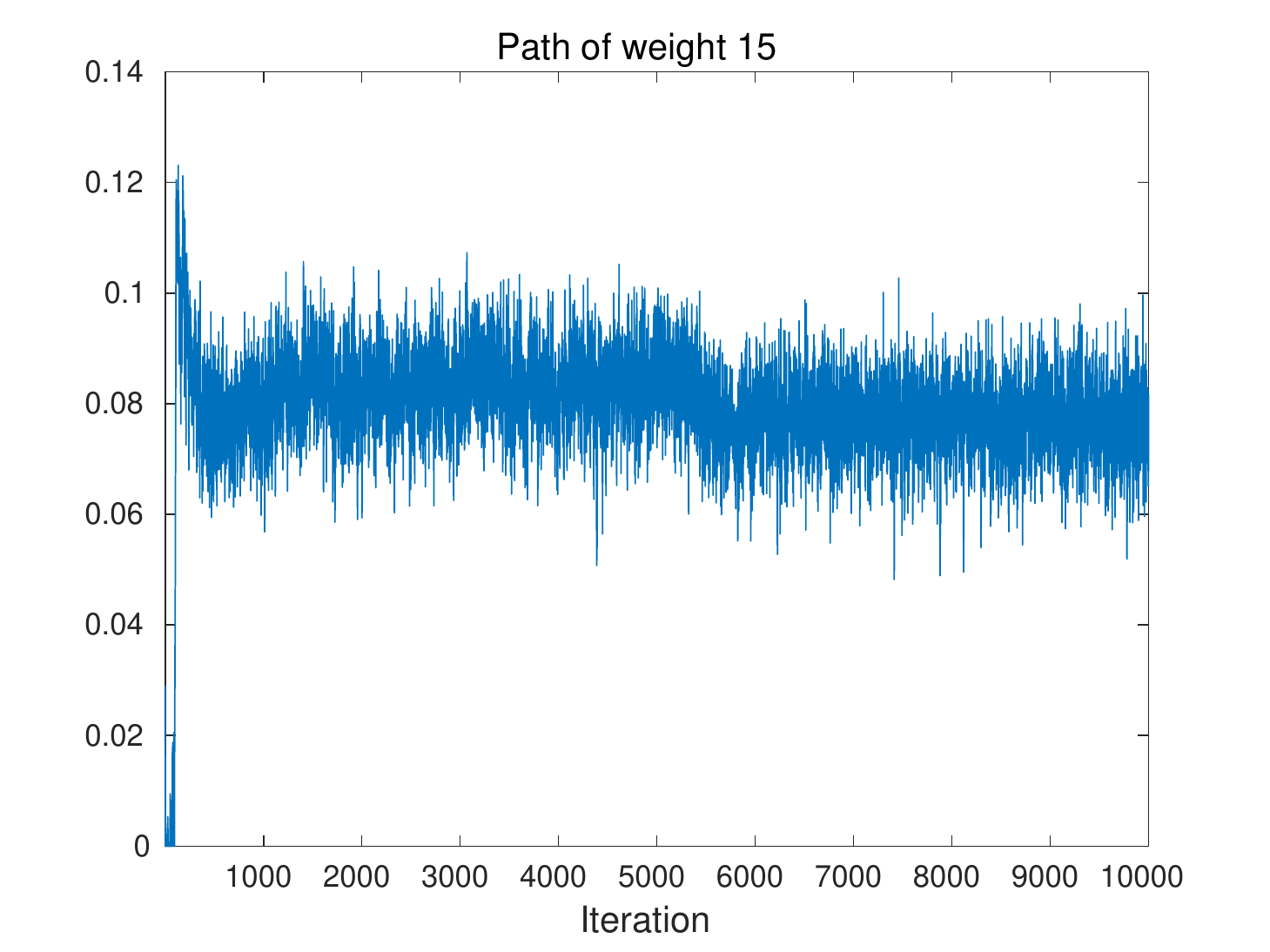}}}
\captionsetup{font=small}
\caption{One of the paths produced by the projected GLD algorithm for the weight of: asset 1 (a) asset 15 (b).}
\label{weightPath}
\end{center}
\end{figure}

For the 15 asset case, the output from the GLD algorithm is a portfolio with equal weights in 14 of the assets and zero weight in one asset. As argued above, this represents a global optimum for the 15 asset problem. As a comparison, a local solver was started from the generated starting points $\wv_0$ for each of the $n_{sim}$ outer simulations of the GLD algorithm. The built-in interior point solver in MATLAB was used as local solver. For all of the generated starting points, the output from the local solver is the equally weighted portfolio with non-zero weights in all of the 15 assets. Thus, the multistart projected GLD algorithm jumps between the local optima and is able to locate the global optimum for the 15 asset problem, whereas a multistart algorithm which uses a local solver finds a local optimum in all cases. The distribution of the final iterate from the GLD algorithm for one of the assets is illustrated in Figure \ref{weightDist}. Figure \ref{weightDist} (a) displays the full distribution where many of the final iterates are concentrated around zero, whereas Figure \ref{weightDist} (b) shows the distribution zoomed in around the non-zero weights. The ability of the GLD algorithm to produce iterates that jump between different local optima is illustrated by the graphs in Figure \ref{weightPath}. In general it is not possible to verify if the output from the GLD algorithm is a global optimum, but the experiments indicate that the algorithm is a useful tool for locating the global optimum for problems for which the number of assets is out of reach for the BB algorithm.

\section{Conclusions}
In this paper we have introduced a portfolio diversification framework based on a novel measure called portfolio dimensionality. This measure is directly related to the tail risk of the portfolio and it is leverage invariant, which means that it can typically be expressed as the ratio of convex functions. In order to solve the global optimization problem that arises when maximizing portfolio dimensionality, two complementary global optimization algorithms have been formulated, one deterministic BB algorithm and one stochastic GLD algorithm. Solving the problem with the BB algorithm, one can guarantee that the global optimum has been found. However, it suffers from the curse of dimensionality which limits the size of the problem when the BB algorithm is used for the optimization. A complementary stochastic optimization algorithm for the global optimization problem has therefore been formulated. As illustrated in Section \ref{Sec4_3}, the multistart projected GLD algorithm can find the global optimum for cases when a multistart local solver algorithm does not. The projected GLD algorithm therefore complements the BB algorithm and allows for solving problems in higher dimensions, albeit without the guarantee that the global optimum will be found. An alternative solution method could be to run the BB algorithm for a fixed number of iterations when solving larger problems. Empirically we observed that the vast majority of the solution time for the BB algorithm is spent on proving optimality for a point found early on. This is a heuristic method that may be used instead of the projected GLD algorithm when solving larger problems. Furthermore, we observed empirically that for problem instances where all correlations are positive, a local solver finds the global optimum as verified by the BB algorithm. This observation may be an indication that when all correlations are positive, the problem is quasi-convex when formulated as a minimization problem.

Our introduced framework extends the diversification frameworks in the literature that are based on only the covariance matrix. Through numerical experiments we have illustrated that our framework possess desirable properties as introduced in the portfolio diversification literature. This can be contrasted to commonly used diversification frameworks such as risk parity and the most diversified portfolio. In order to avoid the problem of obtaining robust estimates of asymmetric tail dependencies between asset returns \citep[see][]{Fra2005}, we have in this paper chosen to model the dependence structure with a Gaussian copula. It is possible to extend the framework to also taking dynamic volatilities and correlations as well as non-linear dependence into account. The model can be extended by using a dynamic GARCH model with skewed and leptokurtic innovations for the marginal distributions, as well as a dynamic conditional correlation model for the copula correlations, see \cite{Eng2002}. Furthermore, in order to capture the asymmetric tail dependence observed in the financial markets, a skewed t copula can be used as in \cite{Chri2012}. Alternatively, a non-linear dependence structure can be modelled with regime shifts as in \cite{Ang2002}.

\section{Acknowledgements}
This work was supported by two EPSRC Impact Acceleration Account grants  [EPSRC IAA PIII011 and EPSRC P049].

\bibliographystyle{rQUF}
\bibliography{diversificationRef}

\begin{appendices}
\section{Useful notation for higher order portfolio moments}\label{app1}
Given the weight vector $\wv\in \mathbb{R}^{n\times 1}$ of relative portfolio weights and the random vector $\rv\in\mathbb{R}^{n\times 1}$ containing the returns of the assets in the portfolio, the third and fourth central moments of the portfolio return $r_p$ are given by
\begin{align}
  \mu_3=\E((r_p-\mu_p)^3)=\E((\wv^{\top}(\rv-\mv))^3)&=\wv^{\top}\Mt(\wv\otimes\wv), \text{ and}\label{mom3_eq} \nlt
  \mu_4=\E((r_p-\mu_p)^4)=\E((\wv^{\top}(\rv-\mv))^4)&=\wv^{\top}\Mf(\wv\otimes\wv\otimes\wv), \label{mom4_eq}
\end{align}
where $\Mt$ and $\Mf$ denote the third and fourth co-moment matrices, respectively, $\otimes$ denotes the Kronecker product, $\mu_p=\E(r_p)$ and $\mv=\E(\rv)$. In reality, the third and fourth co-moments of the asset returns are three and four dimensional tensors with dimensions $n\times n \times n$ and $n\times n \times n \times n$, respectively. In order to make these tensors mathematically tractable we follow \cite{Ath2003} and \cite{Jon2006}, among others, and convert these tensors into two-dimensional matrices. The $n \times n^2$ third co-moment matrix $\Mt$ in (\ref{mom3_eq}) is defined as
\begin{align}
  \Mt=\E\left((\rv-\mv)(\rv-\mv)^{\top} \otimes (\rv-\mv)^{\top} \right)=\{s_{ijk} \},
\end{align}
where
\begin{align}\label{s_eq}
  s_{ijk}=\E\left((r_i-\mu_i)(r_j-\mu_j)(r_k-\mu_k) \right), \sp i,j,k=1,\hdots, n.
\end{align}
The third co-moment matrix can be written in the following block matrix form
\begin{align}\label{Meq2}
  \Mt=\left[\Sm_{1jk} \sp \Sm_{2jk} \sp \hdots \sp \Sm_{njk}  \right],
\end{align}
where the $n \times n$ matrices $\Sm_{ijk}$ are given by
\begin{equation}
  \Sm_{ijk}=
\left[
\begin{array}{c c c c}
s_{i11} & s_{i12}&  \hdots & s_{i1n}\\
s_{i21} & s_{i22} & \hdots & s_{i2n} \\
\vdots & \vdots & \vdots & \vdots \\
s_{in1} & s_{in2} & \hdots & s_{inn}
\end{array}
\right], \sp i=1,\hdots n.
\end{equation}
Similarly, the $n \times n^3$ fourth co-moment matrix $\Mf$ in (\ref{mom4_eq}) is defined as
\begin{align}
  \Mf=\E\left((\rv-\mv)(\rv-\mv)^{\top} \otimes (\rv-\mv)^{\top}\otimes (\rv-\mv)^{\top} \right)=\{k_{ijkl} \},
\end{align}
where
\begin{align}\label{k_eq}
  k_{ijkl}=\E\left((r_i-\mu_i)(r_j-\mu_j)(r_k-\mu_k)(r_l-\mu_l) \right), \sp i,j,k,l=1,\hdots, n,
\end{align}
and the block matrix form is given by
\begin{align}\label{Meq2}
  \Mf=\left[\Km_{11kl} \sp \Km_{12kl} \sp \hdots \sp \Km_{1nkl} \sp | \sp \Km_{21kl} \sp \hdots \sp \Km_{2nkl} \sp | \sp \hdots \sp| \sp \Km_{n1kl} \sp \hdots \sp \Km_{nnkl}  \right],
\end{align}
where the $n \times n$ matrices $\Km_{ijkl}$ are given by
\begin{equation}
  \Km_{ijkl}=
\left[
\begin{array}{c c c c }
k_{ij11} & k_{ij12}&  \hdots & k_{ij1n}\\
k_{ij21} & k_{ij22} & \hdots & k_{ij2n} \\
\vdots & \vdots & \vdots & \vdots \\
k_{ijn1} & k_{ijn2} & \hdots & k_{ijnn}
\end{array}
\right], \sp i,j=1,\hdots n.
\end{equation}
The dimension of each block in (\ref{Meq2}) is given by
\begin{align}
  \left[\Km_{i1kl} \sp \Km_{i2kl} \sp \hdots \sp \Km_{inkl} \right]\in\mathbb{R}^{n \times n^2}, \sp i=1, \hdots, n,
\end{align}
and hence $\Mf\in \mathbb{R}^{n \times n^3}$. As is apparent from equations (\ref{s_eq}) and (\ref{k_eq}), the matrices $\Mt$ and $\Mf$ contain certain symmetries, which means that not all elements need to be explicitly computed. The number of unique elements in $\Mt$ is $n(n+1)(n+2)/6$ and $\Mf$ contains $n(n+1)(n+2)(n+3)/24$ unique elements \citep[see e.g.][]{Jon2006}. The number of unique elements in $\Mt$ and $\Mf$, respectively, for different portfolio sizes are summarized in Table \ref{uniqueelem}.
\begin{table}[h]
\begin{center}
  \begin{tabular}{ c  c  c c c c c}
    \hline \hline
    Portfolio size & 2 & 3 & 4 & 10 & 50 & 100 \\
    \hline
    Number of unique elements in $\Mt$ & 4 & 10 & 20 & 220 & 22,100 & 171,700\\
    Number of unique elements in $\Mf$ & 5 & 15 & 35 & 715 & 292,825 & 4,421,275\\
        \hline
  \end{tabular}
\end{center}
\captionsetup{font=small}
\caption{Number of unique elements in $\Mt$ and $\Mf$ for different portfolio sizes.}
\label{uniqueelem}
\end{table}
As is apparent from the table the number of unique elements grows dramatically with portfolio size. This leads to large estimation errors when attempting to estimate the co-moments from historical return data. In the literature this curse of dimensionality is typically handled by assuming that the asset returns are generated by a factor model through which the number of parameters to be estimated is considerably reduced. Examples from the literature are \cite{Mar2010}, who use a single factor model, and \cite{Bou2015}, who use a multi-factor model.

The third and fourth central moments of the portfolio return are homogeneous functions of degree three and four, respectively. Their gradients can be found by applying Euler's theorem for positively homogeneous functions
\begin{align}
  &\wv^{\T}\nw \mu_3=3\mu_3=\wv^{\T} 3\Mt(\wv\otimes\wv), \text{ and} \nlt
  &\wv^{\T}\nw \mu_4=4\mu_4=\wv^{\T} 4\Mf(\wv\otimes\wv\otimes\wv),
\end{align}
and hence
\begin{align}
  \nw \mu_3= 3\Mt(\wv\otimes\wv), \text{ and} \nlt
  \nw \mu_4= 4\Mf(\wv\otimes\wv\otimes\wv),
\end{align}
where each component of the gradients are homogeneous functions of degree two and three, respectively. Letting $f_i$ denote the $i$th component of $\nw \mu_3$ and applying Euler's theorem for homogeneous functions yields
\begin{align}
  \nw f_i \wv^{\T}=2 f_i.
\end{align}
Letting $J$ denote the Jacobian, the Hessian of $\mu_3$ is given by
\begin{align}
  \nw^2 \mu3=\left( J(\nw \mu_3)\right)^{\T}=\left[\dfrac{\partial \nw \mu_3}{\partial w_1} \sp \dfrac{\partial \nw \mu_3}{\partial w_2} \sp \hdots \sp \dfrac{\partial \nw \mu_3}{\partial w_n}\right]^{\top}=\left[\nw f_1 \sp \nw f_2 \sp \hdots \sp \nw f_n \right].
\end{align}
Since the Hessian is symmetric one obtains
\begin{align}
  \nw^2 \mu_3 \wv=\left[\begin{array}{c}
 (\nw f_1)^{\T} \\ (\nw f_2)^{\T} \\ \vdots \\ (\nw f_n)^{\T} \end{array}\right]\wv=2 \left[\begin{array}{c}
 f_1 \\ f_2 \\ \vdots \\ f_n \end{array}\right] = 2\nw \mu3,
 \end{align}
 and thus
 \begin{align}
   \nw^2 \mu_3=6 \Mt( \wv \otimes \In),
 \end{align}
 where $\In$ denotes the $n \times n$ identity matrix. Similarly, the Hessian of $\mu_4$ is given by
 \begin{align}
   \nw^2 \mu_4=12 \Mf(\wv \otimes \wv \otimes \In).
 \end{align}
\section{Generating the return distribution}\label{simsec}
In this section we describe a meta-Gaussian distribution \citep[see][]{McN2005} defined as the combination of the Gaussian copula and arbitrary distributions for the margins. The meta-Gaussian distribution is in this paper used to generate sample returns from a multivariate distribution with a desired linear dependence structure and marginal distributions with arbitrary skewness and kurtosis parameters. The Normal Inverse Gaussian (NIG) distribution is used for generating marginal distributions with possibly different skewness and kurtosis parameters.
\subsubsection*{The NIG distribution}
The NIG distribution is a special case of the Generalized Hyperbolic distribution introduced by \cite{Bar1977}. It was introduced by \cite{Bar1997} and is commonly used in financial applications to model skewed and leptokurtic distributions. The univariate probability density function (PDF) of the NIG distribution can be expressed as
\begin{align}
  f_X(x)=\dfrac{\delta \alpha \exp\left(\delta \sqrt{\alpha^2-\beta^2} \right)K_1\left(\alpha\sqrt{\delta^2+(x-\mu)^2} \right)}{\pi\sqrt{\delta^2+(x-\mu)^2}},
\end{align}
where $\delta>0$, $0\leq |\beta|< \alpha$, and $K_1$ is the modified Bessel function of the third kind of order 1, see \cite{Abr1972}. The parameters $\mu$ and $\delta$ determine the location and scale, respectively, while $\alpha$ and $\beta$ control the shape of the density. In particular, $\beta=0$ corresponds to a symmetric distribution. The mean, variance, skewness and kurtosis of the NIG distribution are given by
\begin{align}
  &\E(X)=\mu+\dfrac{\delta \beta}{\sqrt{\alpha^2-\beta^2}}, \, \text{Var}(X)=\dfrac{\delta\alpha^2}{(\alpha^2-\beta^2)^{3/2}}, \, \gamma=3\dfrac{\beta}{\alpha}\dfrac{1}{\delta^{1/2}(\alpha^2-\beta^2)^{1/4}}, \mbox{ and } \nl
  &\kappa=3+3\left(1+4\left(\dfrac{\beta}{\alpha} \right)^2 \right)\dfrac{1}{\delta(\alpha^2-\beta^2)^{1/2}}.
\end{align}
From the condition $0\leq |\beta|<\alpha$, it follows that the skewness-kurtosis bound $\gamma^2<3(\kappa-3)/5$, must hold for the NIG distribution.
\subsubsection*{The Gaussian copula}
Sklar's Theorem \citep{Sklar1959} shows that all multivariate cumulative distribution functions (CDFs) contain copulas and that copulas may be used together with univariate CDFs in order to construct multivariate CDFs. A formulation of Sklar's theorem, taken from \cite{McN2005}, is given below.
\begin{theorem}[Sklar 1959] Let F be a joint CDF with marginal CDFs $F_1,\hdots,F_d$. Then there exists a copula $C: [0,1]^d \rightarrow [0,1]$ such that, for all $x_1,\hdots,x_d \in (-\infty,\infty)$,
\begin{align}\label{copula}
  F(x_1,\hdots,x_d)=C(F_1(x_1), \hdots, F_d(x_d)).
\end{align}
If the margins are continuous, then $C$ is unique. Conversely, if $C$ is a copula and $F_1, \hdots, F_d$ are univariate CDFs, then the function $F$ defined in (\ref{copula}) is a joint CDF with margins $F_1, \hdots, F_d$.
\end{theorem}
\noindent Thus, modelling the multivariate return distribution with copulas allows for separating the modelling of the dependence structure and the marginal distributions. In particular, it allows for modelling of the marginal asset return distributions with differing skewness and kurtosis. In this paper, the Gaussian copula is used for modelling of the dependence structure between the asset returns. Let $\boldsymbol{\varPhi}_{\mathbf{R}}$ denote the joint CDF of the $d$-dimensional normally distributed $\Xv \sim N_d(\mathbf{0},\mathbf{R})$, where $\mathbf{R}\in \R^{d\times d}$ denotes the correlation matrix. The Gaussian copula is then given by
\begin{align}
  C_{\mathbf{R}}^{\text{Ga}}(u_1,\hdots,u_d)=\boldsymbol{\varPhi}_{\mathbf{R}}\left(\varPhi^{-1}(u_1),\hdots,\varPhi^{-1}(u_d) \right),
\end{align}
were $\varPhi$ denotes the standard univariate Gaussian CDF.
\subsubsection*{Adjusting the input correlation matrix}
When generating simulated returns from the meta-Gaussian distribution, one needs to take into account that the realized correlation matrix of the generated returns depends on the copula as well as the marginal return distributions. From Hoeffding's covariance identity \citep{Hoe1940}, the covariance between two random variables $X$ and $Y$ can be expressed as
\begin{align}\label{Hoeffding}
  \text{Cov}(X,Y)=\int\limits_{-\infty}^{\infty}\int\limits_{-\infty}^{\infty}\left(F_{X,Y}(x,y)-F_X(x)F_Y(y) \right)dx dy,
\end{align}
where $F_{X,Y}(\cdot,\cdot)$ denotes the joint CDF and $F_X(\cdot), F_Y(\cdot)$ denote the two marginal CDFs. Using Sklar's theorem, equation (\ref{Hoeffding}) can be expressed as
\begin{align}
  \text{Cov}(X,Y)=\int\limits_{-\infty}^{\infty}\int\limits_{-\infty}^{\infty}\left(C(F_X(x),F_Y(y))-F_X(x)F_Y(y) \right)dx dy,
\end{align}
where $C$ for the meta-Gaussian distribution is given by the Gaussian copula. Let $C_{\rho_{\text{in}}}^{\text{Ga}}$ denote the two-dimensional Gaussian copula with linear correlation parameter $\rho_{\text{in}}$. The correlation between the two variables $X$ and $Y$ as a function of the correlation parameter $\rho_{\text{in}}$ is then given by
\begin{align}\label{corrInv}
  \rho_{\text{out}}(\rho_{\text{in}})=\dfrac{\int\limits_{-\infty}^{\infty}\int\limits_{-\infty}^{\infty}\left(C_{\rho_{\text{in}}}^{\text{Ga}}(F_X(x),F_Y(y))-F_X(x)F_Y(y) \right)dx dy}{\sqrt{\text{Var(X)}{\text{Var(Y)}}}}.
\end{align}
For the generation of samples with the desired realized linear correlations between returns, the input correlation matrix for the Gaussian copula is used by inverting (\ref{corrInv}). Its entries are determined by numerical integration as, in general, no analytic solution for the inverse exists.
\subsubsection*{Generating returns from the meta-Gaussian distribution}
The procedure for generating returns from the meta-Gaussian distribution is given below.
\begin{itemize}
  \item[(1)] Given the marginal distributions and the Gaussian copula, find the input correlation matrix $\mathbf{R}_{\text{in}}$ by numerical inversion of equation (\ref{corrInv}).
  \item[(2)] Generate $\boldsymbol{Z}\in N_d\left(\boldsymbol{0},\mathbf{R}_{\text{in}}\right)$.
  \item[(3)] Calculate $\boldsymbol{U} =\left[\varPhi(Z_1), \hdots, \varPhi(Z_d) \right]^{\T}$. The CDF of the random vector $\boldsymbol{U}$ is given by $C_{\mathbf{R}_{\text{in}}}^{\text{Ga}}$.
  \item[(4)] Use quantile transformation to obtain $\Xv=\left[F_1^{-1}(U_1), \hdots, F_d^{-1}(U_d) \right]^{\T}$ by numerical inversion of the marginal CDFs. The random vector $\Xv$ has marginal CDFs $F_1,\hdots,F_d$ and multivariate CDF $C_{\mathbf{R}_{\text{in}}}^{\text{Ga}}\left( F_1(x_1), \hdots, F_d(x_d)\right)$.
  \end{itemize}

  \end{appendices}

\end{document}